\documentclass[floatfix,%
 reprint,%
 amssymb, amsmath,%
 aip,jcp,
frontmatterverbose,
]{revtex4-1}

\usepackage{amsmath,amssymb,amsfonts,mathtools}
\usepackage{multirow}
\usepackage{float}
\usepackage{csquotes}
\usepackage{dsfont}
\usepackage[mathscr]{euscript}

\usepackage{graphicx}
\usepackage{bm}%
\usepackage{longtable}
\usepackage{supertabular}
\usepackage{rotating}
\usepackage[colorlinks=false,linkcolor=blue]{hyperref}
\usepackage[normalem]{ulem}
\usepackage{ifthen}
\newboolean{includeMQDT}
\setboolean{includeMQDT}{false}  

\usepackage[dvipsnames]{xcolor}

\usepackage{dcolumn,ulem}  
\usepackage{xspace}

\begin{document}
%
%
%
\title{\large Non-Adiabatic Ring Polymer Molecular Dynamics with Spin Mapping Variables}
\author{Duncan Bossion}
\email{dbossion@ur.rochester.edu}
\affiliation{Department of Chemistry, University of Rochester, 120 Trustee Road, Rochester, New York 14627, USA}
\author{Sutirtha N. Chowdhury}
\affiliation{Department of Chemistry, University of Rochester, 120 Trustee Road, Rochester, New York 14627, USA}
\author{Pengfei Huo}
\email{pengfei.huo@rochester.edu}
\affiliation{Department of Chemistry, University of Rochester, 120 Trustee Road, Rochester, New York 14627, USA}
\affiliation{The Institute of Optics, Hajim School of Engineering, University of Rochester, Rochester, New York, 14627}
%
\begin{abstract}
We present a new non-adiabatic ring polymer molecular dynamics (NRPMD) method based on the spin mapping formalism, which we refer to as the spin-mapping NRPMD (SM-NRPMD) approach. We derive the path-integral partition function expression using the spin coherent state basis for the electronic states and the ring polymer formalism for the nuclear degrees of freedom (DOFs). This partition function provides an efficient sampling of the quantum statistics. Using the basic property of the Stratonovich-Weyl transformation, we derive a Hamiltonian which we propose for the dynamical propagation of the coupled spin mapping variables and the nuclear ring polymer. The accuracy of the SM-NRPMD method is numerically demonstrated by computing nuclear position and population auto-correlation functions of non-adiabatic model systems. The results from SM-NRPMD agree very well with the numerically exact results. The main advantage of using the spin mapping variables over the harmonic oscillator mapping variables is numerically demonstrated, where the former provides nearly time-independent expectation values of physical observables for systems under thermal equilibrium, the latter can not preserve the initial quantum Boltzmann distribution. We also explicitly demonstrate that SM-NRPMD provides invariant dynamics upon various ways of partitioning the state-dependent and state-independent potentials.
\end{abstract}
\maketitle
\section{Introduction}
One of the central challenges in theoretical chemistry is to accurately simulate chemical reactions involving non-adiabatic processes and nuclear quantum effects.\cite{tully2012} These reactions, such as the electron transfer, proton-coupled electron transfer, or the scattering reactions involving non-adiabatic transitions among many electronic states and nuclear quantum effects, are commonly encountered from photo-catalysis, biochemistry and enzymatic reactions, to astrochemistry. Developing accurately yet numerically efficient approaches became a key focus in physical chemistry.

To this end, a large number of these approaches are developed, including the popular trajectory surface-hopping method (mixed quantum-classical approach),\cite{tully1990,subotnik2016,wang2016,Barbatti} the Linearized semi-classical (LSC) path-integral approaches,\cite{miller2001,miller2009} partially linearized density matrix (PLDM) path-integral approaches,\cite{huo2011,huo2011molphys,huo2013,lee2016} the mixed quantum-classical Liouville equation,\cite{kim2008,kapral2010,kelly2012,Hsieh2013} and the symmetrical quasi-classical (SQC) approach,\cite{Miller:2016b,cotton2019} 
to name a few. Despite their successes, these approaches generally do not preserve quantum detailed balance\cite{parandekar2005,schmidt2008} or zero-point energy (ZPE) associated with the nuclear degrees of freedom (DOFs), and often suffer from numerical issues such as ZPE leakage.\cite{habershon2009,muller1999}

Imaginary-time path-integral approaches,\cite{berne1986,ceperley1995,chandler1981} such as the ring polymer molecular dynamics (RPMD),\cite{habershon2013,craig2004} resemble classical dynamics in an extended phase space and provide a convenient way to compute approximate quantum time-correlation functions.\cite{habershon2013} The classical evolution of RPMD preserves its initial quantum distribution captured by the ring polymer Hamiltonian, and it is free of the zero-point energy leaking problem.\cite{habershon2009,habershon2013} Despite its success in describing quantum effects in the condensed phase, RPMD is limited to one-electron non-adiabatic dynamics\cite{menzeelev2010,menzeelev2011,kretchmer2013,kretchmer2016,Ananth2016} or nuclear quantization,\cite{habershon2013,boekelheide2011,markland2014,Markland:2017,Ceriotti:2018} as well as the lack of real-time electronic coherence effects.\cite{menzeelev2010,menzeelev2011} 

Recently emerged state-dependent RPMD approaches, such as non-adiabatic RPMD (NRPMD),\cite{richardson2013,richardson2017,chowdhury2019} mapping variable RPMD (MV-RPMD),\cite{ananth2013,ananth2017} and  coherent state RPMD (CS-RPMD)\cite{chowdhury2017} are promising to provide accurate non-adiabatic dynamics with an explicit description of electronic states, in addition to the reliable treatment of nuclear quantum effects through ring polymer quantization. The common ingredient of these approaches is the Meyer-Miller-Thoss-Stock (MMST) mapping formalism,\cite{meyer1979_2,stock1997,thoss1999} which maps $N$ electronic states onto $N$ singly excited harmonic oscillators. The electronic non-adiabatic dynamics are hence mapped onto the phase space trajectories of the mapping oscillators, which evolve together with the nuclear ring polymer. Hence, these MMST-based RPMD approaches can be viewed as unified theories of the mapping oscillators and the ring polymer. These methods are shown to provide both accurate non-adiabatic dynamics as well as nuclear quantum effects.\cite{richardson2013,richardson2017,chowdhury2019} In particular, the NRPMD approach has been rigorously derived from the non-adiabatic Matsubara dynamics framework.\cite{chowdhury2021}

One potential limitation of these state-dependent RPMD approaches is rooted in the MMST mapping representation. It is well known that the MMST representation has a larger size of Hilbert space than the original electronic subspace, and requires projection back to that subspace to obtain accurate results.\cite{ananth2010,kelly2012} In addition, the total population along a single trajectory is not guaranteed to be unitary, hence breaking the dynamical invariance under different ways of partitioning the potentials into the state-dependent and state-independent components.\cite{thoss1999,kelly2012} Besides the widely used MMST representation, there exist other mapping formalisms based upon spin coherent states.\cite{klauder1979,meyer1979,lucke1999,garg2003,song2006} In particular, a new spin mapping formalism\cite{richardson2019,richardson2020} based on the Stratonovich-Weyl transform\cite{klimov} was recently developed by Runeson and Richardson. In this spin-mapping approach,\cite{richardson2019} two electronic states are mapped onto two angles defining the spin coherent state on the Bloch sphere. One of the advantages of this approach, compared to the MMST formalism, is that the dimensionality of the spin coherent state basis is of the same size of the electronic states of the original system, hence it provides a more consistent mapping than the MMST approach and it does not require additional projections back to the electronic subspace.\cite{richardson2019} The spin mapping (SM) variables, being bounded on the Bloch sphere, also  guarantees the total population along a single trajectory to be unitary. This further enforces the independence of the dynamics to the splitting between the state-dependent and state-independent parts of the Hamiltonian. It has been shown that in the LSC\cite{richardson2019, richardson2020} and the PLDM\cite{mannouch2020,mannouch2020_2} approaches, using spin-mapping approach provides more accurate non-adiabatic dynamics compared to the corresponding approaches when using the MMST formalism.\cite{miller2009, huo2011} These exciting theoretical developments of the spin mapping variables motivate us to develop the NRPMD approach with the spin mapping representation. 

In this paper, we develop a new non-adiabatic RPMD method which we refer to as the spin mapping NRPMD (SM-NRPMD) approach, based on the recently developed spin mapping formalism.\cite{richardson2019,richardson2020} We first derive a partition function formalism based on the SM representation that allows one to efficiently sample the exact quantum statistics. We then derive a SM-NRPMD Hamiltonian for propagating dynamics. With the proposed SM-NRPMD approach, we compute the Kubo-transformed position and population auto-correlation functions with non-adiabatic model systems, and demonstrate that this approach is capable of accurately describing both the correct quantum statistics as well as the electronic Rabi oscillations. Compared to the MMST-based NRPMD approaches,\cite{richardson2020rev} SM-NRPMD seems to preserve the quantum detailed balance, resulting in a nearly time-independent expectation value of the nuclear position or population for the system under the thermal equilibrium. Finally, we demonstrate that the dynamics is invariant of the partitioning of the potential into the state-dependent and the state-independent components.

\section{Basic Theory of the Spin Mapping Formalism}
In this section, we review the spin-mapping for electronic states introduced by Runeson and Richardson.\cite{richardson2019, richardson2020} A comprehensive introduction of this material can be found in Ref.~\citenum{richardson2019}.

The total Hamiltonian operator of the system is
\begin{align}
\hat{H} = \frac{\hat{P}^2}{2m}\hat{\mathcal{I}} + U_0(\hat{R})\hat{\mathcal{I}} + \begin{pmatrix}
V_1(\hat{R}) & \Delta(\hat{R})\\ 
\Delta(\hat{R}) & V_2(\hat{R})
\end{pmatrix},
\end{align}
where $U_0(\hat{R})$ represents the state-independent potential energy operator and $\hat{R}$ and $\hat{P}$ are the position and momentum operators of the nuclear degrees of freedom (DOFs), respectively. The Hamiltonian can also be written in terms of the spin operator as\cite{meyer1979}
\begin{equation}\label{eq:spin_ham}
\hat{H} = H_0\hat{\mathcal{I}} + \frac{1}{\hbar}\mathbf{H}\cdot\hat{\mathbf{S}}=H_0\hat{\mathcal{I}} + \frac{1}{\hbar}({H}_{x}\cdot\hat{S}_{x}+ {H}_{y}\cdot\hat{S}_{y}+ {H}_{z}\cdot\hat{S}_{z}),
\end{equation}
where $\hat{\mathcal{I}}$ is the $2\times 2$ identity matrix, $\hat{S}_i = \frac{\hbar}{2}\hat{\sigma}_{i}~(\mathrm{for} \ i\in\{x,y,z\})$ is the quantum spin operator, with $\hat{\sigma}_i$ being the Pauli matrices expressed as follows
\begin{align}
\hat{\sigma}_x = \begin{pmatrix}
0 & 1 \\ 
1 & 0
\end{pmatrix}, & \hspace{0.2cm} \hat{\sigma}_y = \begin{pmatrix}
0 & -i \\ 
i & 0
\end{pmatrix}, \hspace{0.2cm} \hat{\sigma}_z = \begin{pmatrix}
1 & 0 \\ 
0 & -1
\end{pmatrix}.
\end{align}
Various components of the Hamiltonian in Eq.~\ref{eq:spin_ham} are expressed as
\begin{subequations}
\begin{align}
H_0 &= \frac{\hat{P}^2}{2m} + U_0(\hat{R}) + \frac{1}{2}(V_1(\hat{R}) + V_2(\hat{R})), \\
H_x &= 2 \Re(\Delta(\hat{R})), \label{Hx}\\
H_y &= 2 \Im(\Delta(\hat{R})), \label{Hy}\\
H_z &= V_1(\hat{R}) - V_2(\hat{R}),\label{Hz}
\end{align}
\end{subequations}
where $\Re$ and $\Im$ represents the real and imaginary part of an operator, respectively. Note that for a molecular Hamiltonian, one often has $\Im(\Delta(\hat{R}))=0$.

Following the original work on spin-mapping variables,\cite{richardson2019} we introduce the spin coherent state (SCS) basis\cite{radcliffe1971,richardson2019}
\begin{align}
|\mathcal{\bf u}\rangle = \cos\frac{\theta}{2}e^{-i\varphi/2}|1\rangle + \sin\frac{\theta}{2}e^{i\varphi/2}|2\rangle,
\end{align}
with the two angles, $\theta$ and $\varphi$, defining the state of spin of the system on the Bloch sphere. The SCS vector is normalized $\langle\mathbf{u}|\mathbf{u}\rangle= 1$. The expectation value of the spin operator is 
\begin{equation}\label{eq:spin-exp}
S_{i}(\mathbf{u})=\langle \mathbf{u}|\hat{S}_{i}|\mathbf{u}\rangle=\frac{\hbar}{2}u_i, \hspace{0.5cm} i\in\{x,y,z\},
\end{equation}
where $u_x$, $u_y$, and $u_z$ are expressed as follows
\begin{subequations}
\begin{align}
u_x &= \sin\theta\cos\varphi, \\
u_y &= \sin\theta\sin\varphi, \\
u_z &= \cos\theta.
\end{align}
\end{subequations}

We further introduce three different functions to define for the \textit{Stratonovich-Weyl} (SW) transformation of any operator in the SM representation and hence obtain expectation values. They are the Q-, P- and W-functions. These functions depend on the \emph{kernel} $\hat{w}_\mathrm{s}$ and the \emph{spin radius} $r_\mathrm{s}$ as follows\cite{richardson2019}
\begin{subequations}
\begin{align}
\hat{w}_\mathrm{s}(\mathbf{u})&= \frac{1}{2}\hat{\mathcal{I}} + r_\mathrm{s}\mathbf{u}\cdot\hat{\boldsymbol{\sigma}}, \hspace{0.5cm} \mathrm{s}\in\{\mathrm{Q},\mathrm{P},\mathrm{W\}}, \label{ws}\\
r_\mathrm{Q} &= \frac{1}{2}, \hspace{0.2cm} r_\mathrm{P} = \frac{3}{2}, \hspace{0.2cm} r_\mathrm{W} = \frac{\sqrt{3}}{2},
\end{align}
\end{subequations}
where $\mathbf{u}\cdot\hat{\boldsymbol{\sigma}}=u_x\cdot\hat{\sigma}_x+u_y\cdot\hat{\sigma}_y+u_z\cdot\hat{\sigma}_z$.

The SCS projection operator is $|\mathbf{u}\rangle\langle\mathbf{u}|=\cos^2\frac{\theta}{2}|1\rangle\langle 1|+\cos\frac{\theta}{2}\sin\frac{\theta}{2}e^{-i\varphi}|1\rangle \langle 2|+\cos\frac{\theta}{2}\sin\frac{\theta}{2}e^{i\varphi}|2\rangle \langle 1|+\sin^2\frac{\theta}{2}|2\rangle\langle 2|$. Note that $|\mathbf{u}\rangle\langle\mathbf{u}|$  is equivalent to $\hat{w}_\mathrm{Q}$ as
\begin{equation}\label{wq-proj}
\hat{w}_\mathrm{Q}=|\mathbf{u}\rangle\langle\mathbf{u}|,
\end{equation}
which can be easily verified using  elementary trigonometric identities. On the other hand, $\hat{w}_\mathrm{P}$ and $\hat{w}_\mathrm{W}$ do not have simple relations with $|\mathbf{u}\rangle\langle\mathbf{u}|$.

\subsection{Spin Mapping of Diabatic Electronic States}
The SW transform of an operator $\hat{A}$ is defined as 
\begin{equation}\label{eq:SW-trans}
A_\mathrm{s}(\mathbf{u}) \equiv [\hat{A}]_\mathrm{s}(\mathbf{u})= \mathrm{Tr}_\mathrm{e}[\hat{A}\hat{w}_\mathrm{s}], 
\end{equation}
where the trace is taken in the electronic subspace, which is equivalent to the 2-state spin subspace.

Mapping an operator $\hat{A}$ onto the spin Hilbert subspace corresponds to the following relation
\cite{richardson2019}
\begin{equation}\label{eq:sfuncdef}
\hat{A}\to A_\mathrm{s}(\mathbf{u})= \mathrm{Tr}_\mathrm{e}[\hat{A}\hat{w}_\mathrm{s}].
\end{equation}
Generalizing the theory to many states is also possible\cite{richardson2020} by using the generators of the
$\mathrm{SU}(N)$ Lie algebra (when $N=3$ it corresponds to the Gell-Mann matrices in the $SU(3)$-symmetry
theory of quarks).

For the $\mathrm{s}=\mathrm{Q}$ special case, this mapping relation means that
\begin{align}
A_\mathrm{Q}(\mathbf{u}) = \mathrm{Tr}_\mathrm{e}[\hat{A}\hat{w}_\mathrm{Q}]=\mathrm{Tr}_\mathrm{e}[\hat{A}|\mathbf{u}\rangle\langle\mathbf{u}|]=\langle\mathbf{u}|\hat{A}|\mathbf{u}\rangle. 
\end{align}
The Q-relation maps the spin operator $\hat{S}_{i}$ with $[\hat{S}_{i}]_\mathrm{Q}=\frac{\hbar}{2}u_{i}$, which is its expectation value in the SCS through Eq.~\ref{eq:spin-exp}. 

Using the spin-mapping defined in Eq.~\ref{eq:sfuncdef}, it is easy to show that $[\hat{\mathcal{I}}]_\mathrm{s}(\mathbf{u})=1$ (because $\mathrm{Tr_e}\hat{\sigma_{i}}=0$ for all $i$), as well as
\begin{equation}
{\bf S}_\mathrm{s}(\mathbf{u})\equiv[\hat{\bf S}]_\mathrm{s}(\mathbf{u})=\mathrm{Tr_e}[\frac{\hbar}{2}\hat{\boldsymbol\sigma}(\frac{1}{2}\hat{\mathcal{I}} + r_\mathrm{s}\mathbf{u}\cdot\hat{\boldsymbol{\sigma}})]=\hbar r_\mathrm{s}{\bf u}.
\end{equation}
The projection operators are transformed as
\begin{subequations}
\begin{align}
&[|1\rangle\langle 1|]_\mathrm{s}(\mathbf{u}) =[\frac{1}{2}\hat{\mathcal{I}}+\frac{1}{\hbar}\hat{S}_{z}]_\mathrm{s}(\mathbf{u})=\frac{1}{2}+r_\mathrm{s}\cos\theta,\label{proj11}\\
&[|2\rangle\langle 2|]_\mathrm{s}(\mathbf{u}) =[\frac{1}{2}\hat{\mathcal{I}}-\frac{1}{\hbar}\hat{S}_{z}]_\mathrm{s}(\mathbf{u})=\frac{1}{2}-r_\mathrm{s}\cos\theta, \label{proj22}\\
&[|1\rangle\langle 2|+|2\rangle\langle 1|]_\mathrm{s}(\mathbf{u}) =2[\frac{1}{\hbar}\hat{S}_{x}]_\mathrm{s}(\mathbf{u})=2r_\mathrm{s}\sin\theta \cos\varphi.
\end{align}
\end{subequations}

The Hamiltonian in Eq.~\ref{eq:spin_ham} is mapped as $\hat{H}\to[\hat{H}]_\mathrm{s}(\mathbf{u})$, with the following expression
\begin{align}\label{eq:sm-ham}
&H_\mathrm{s}(\mathbf{u})\equiv[\hat{H}]_\mathrm{s}(\mathbf{u})= H_{0}+r_\mathrm{s}{\bf H}\cdot \mathbf{u}\\
&=\frac{P^2}{2m} + U_0 +(\frac{1}{2}+r_\mathrm{s}\cos\theta)\cdot V_{1}+(\frac{1}{2}-r_\mathrm{s}\cos\theta)\cdot V_{2}\nonumber\\
&~~~+2r_\mathrm{s}\sin\theta\cos\varphi\cdot\Delta\nonumber.
\end{align}
Note that $H_0$ and $\bf H$ are in principle $R$-dependent. The SW mapping is closely related to the MMST mapping approach, and a brief discussion between these two formalisms is provided in Appendix A, whereas a thorough comparison can be found in Ref.~\citenum{richardson2019}.

To obtain the equations of motion (EOM) governed by ${H}_{\mathrm{s}}(\mathbf{u})$ in Eq.~\ref{eq:sm-ham} for the spin-mapping variables, we start with the following Heisenberg EOM for $\hat{\bf S}$
\begin{equation}
\frac{d}{dt}\hat{\bf S}=\frac{1}{i\hbar}[\hat{\bf S},\hat{H}]=\frac{1}{\hbar}{\bf H}(\hat{R})\wedge\hat{\bf S},
\end{equation}
where $\wedge$ denotes the cross product of two vectors, and we have used the fact that ${H}_0\hat{\mathcal I}$ inside $\hat{H}$ (Eq.~\ref{eq:spin_ham}) is $\hat{\bf S}$-independent, hence commutes with $\hat{\bf S}$. Applying the SW transform (Eq.~\ref{eq:SW-trans}) on both sides of the above equation, we have
\begin{equation}\label{u-dyn}
\frac{d}{dt}{\bf u}=\frac{1}{\hbar}{\bf H}(\hat{R})\wedge{\bf u}.
\end{equation}
Note that the above equation is exact, regardless of the $r_{\mathrm{s}}$-dependence of $\hat{\bf H}$. Of course, the EOM for the nuclear DOF is not yet explicitly expressed. When choosing the Wigner representation for the nuclei and using the quantum-classical Liouville equation (QCLE),\cite{richardson2019} Eq.~\ref{u-dyn} can also be rigorously derived. This equation can be solved by treating $\bf u$ as dynamical variables, or equivalently, $\theta$ and $\varphi$ as dynamical variables. Further analysis of this is provided in Appendix B.

\subsection{Properties of the Stratonovich-Weyl Transform}
Here, we briefly summarize several basic properties of the Stratonovich-Weyl transform, which will be used to derive the quantum partition function and the spin-mapping NRPMD Hamiltonian in the next section. Using the spin mapping, the quantum mechanical trace of an operator $\hat{A}$  in the Q-function is expressed as
\begin{align}\label{Eq:TrA}
\mathrm{Tr}_\mathrm{e}[\hat{A}] &= \int \mathrm{d}\mathbf{u}\langle\mathbf{u}|\hat{A}|\mathbf{u}\rangle = \int \mathrm{d}\mathbf{u}A_\mathrm{Q}(\mathbf{u})\\
&= \frac{1}{2\pi}\int_0^\pi \mathrm{d}\theta\sin\theta\int_0^{2\pi} \mathrm{d}\varphi A_\mathrm{Q}(\theta,\varphi),\nonumber
\end{align}
where $\int \mathrm{d}\mathbf{u}=\frac{1}{2\pi}\int_0^\pi \mathrm{d}\theta\sin\theta\int_0^{2\pi} \mathrm{d}\varphi$.
Note that because $\langle\mathbf{u}|\hat{A}\hat{B}|\mathbf{u}\rangle \neq \langle\mathbf{u}|\hat{A}|\mathbf{u}\rangle\langle\mathbf{u}|\hat{B}|\mathbf{u}\rangle$ (the uncertainty property), the Q-function cannot be used to directly compute the quantum mechanical trace of a product of operators, {\it i.e.}, $\mathrm{Tr_{e}}[\hat{A}\hat{B}]\neq\int d\mathbf{u} A_\mathrm{Q}(\mathbf{u})B_\mathrm{Q}(\mathbf{u})$.

To solve this issue, one can use the P-function and the following property
\begin{equation}
\mathrm{Tr_e}[\hat{A}\hat{B}] = \int \mathrm{d}\mathbf{u}A_\mathrm{Q}(\mathbf{u})B_\mathrm{P}(\mathbf{u}) = \int \mathrm{d}\mathbf{u}A_\mathrm{P}(\mathbf{u})B_\mathrm{Q}(\mathbf{u}).
\label{eq:sfunc1}
\end{equation}
The W-function can also be used for this purpose
\begin{equation}
\mathrm{Tr_e}[\hat{A}\hat{B}] = \int \mathrm{d}\mathbf{u}A_\mathrm{W}(\mathbf{u})B_\mathrm{W}(\mathbf{u}).
\label{eq:sfunc2}
\end{equation}
Summarizing the above properties, we have 
\begin{equation}\label{eq:QPQ}
\mathrm{Tr_e}[\hat{A}\hat{B}] = \int \mathrm{d}\mathbf{u}A_\mathrm{s}(\mathbf{u})B_\mathrm{\bar{s}}(\mathbf{u}),
\end{equation}
where $\{\mathrm{s}$, $\bar{\mathrm{s}}\}$ can be \{Q, P\}, \{P, Q\}, or \{W, W\}. The proof of Eq.~\ref{eq:QPQ} is elementary and is provided in Appendix C.

Choosing $\hat{B}=\hat{\mathcal I}$, Eq.~\ref{eq:QPQ} becomes 
\begin{equation}\label{trA}
\mathrm{Tr_e}[\hat{A}] = \int \mathrm{d}\mathbf{u}A_\mathrm{s}(\mathbf{u})[\hat{\mathcal I}]_\mathrm{\bar{s}}(\mathbf{u})=\int \mathrm{d}\mathbf{u}A_\mathrm{s}(\mathbf{u}),
\end{equation}
where we have used the fact that $[\hat{\mathcal I}]_\mathrm{\bar{s}}=1$. This suggest that for the quantum mechanical trace of an operator $\hat{A}$, one can freely choose $\mathrm{s}\in\{\mathrm{Q,P,W}\}$, which all provide the identical answer, even though different kernel $\hat{w}_\mathrm{s}$ and radius $r_\mathrm{s}$ is used. Further using the definition of $A_\mathrm{s}(\mathbf{u})$ (Eq.~\ref{eq:sfuncdef}) into Eq.~\ref{trA}, we have
\begin{equation}
\mathrm{Tr_e}[\hat{A}]=\int \mathrm{d}\mathbf{u}\mathrm{Tr_{e}}[\hat{A}\hat{w}_\mathrm{s}]=\mathrm{Tr_{e}}\big[\hat{A}\cdot\int \mathrm{d}\mathbf{u}\hat{w}_\mathrm{s}\big],
\end{equation}
where we have moved the $\mathrm{d}\mathbf{u}$ integral inside the trace (and notice that $\hat{A}$ is $\mathbf{u}$ independent). The above equality indicates the following resolution of identity 
\begin{align}\label{resolution}
\mathds{1}_\mathbf{u} &= \int \mathrm{d}\mathbf{u}\hat{w}_\mathrm{s}\\
&=\frac{1}{2\pi}\int_0^\pi \mathrm{d}\theta\sin\theta\int_0^{2\pi} \mathrm{d}\varphi\Big(\frac{1}{2}\hat{\mathcal{I}} + r_\mathrm{s}\mathbf{u}\cdot\hat{\boldsymbol{\sigma}}\Big),\nonumber
\end{align}
where the second line of the above equation used the expressions of $ \int \mathrm{d}\mathbf{u}$ (Eq.~\ref{Eq:TrA}) and $\hat{w}_\mathrm{s}$ (Eq.~\ref{ws}). This identity can also be easily verified through elementary integrals, which is provided in Appendix C. 

When choosing $\mathrm{s=Q}$, the resolution of identity is 
\begin{align}
\mathds{1}_\mathbf{u} &=\int \mathrm{d}\mathbf{u}\hat{w}_\mathrm{Q} =\int \mathrm{d}\mathbf{u}|\mathbf{u}\rangle\langle\mathbf{u}|,
\end{align}
where we used $\hat{w}_\mathrm{Q}=|\mathbf{u}\rangle\langle\mathbf{u}|$ (Eq.~\ref{wq-proj}).


\section{Quantum Partition Function with Spin-Mapping variables}
\subsection{Spin Coherent State (SCS) Partition Function}
The canonical partition function is expressed as ${\cal{Z}} = \mathrm{Tr}_\mathrm{n}\mathrm{Tr}_\mathrm{e}[ e^{-\beta\hat{H}}]$, where $\mathrm{Tr}_\mathrm{n}$ and $\mathrm{Tr}_\mathrm{e}$ represent the trace over the nuclear and electronic DOFs, respectively, and $\beta=1/k_\mathrm{B}T$. The partition function can be exactly evaluated in the limit $N\rightarrow\infty$ by the Trotter discretization,\cite{trotter1959} where $N$ is the number of ring polymer beads. 

We start from expressing the quantum partition function as follows
\begin{align}
{\cal{Z}} = \mathrm{Tr_{e}Tr_{n}}\Big[\big( e^{-\beta_N(H_0\hat{\mathcal I} + \frac{1}{\hbar}\mathbf{H}\cdot\hat{\mathbf{S}})} \big)^N\Big],
\end{align}
where $\beta_N = \beta/N$. Inserting $N$ copies of the identities in the nuclear subspace, $\mathds{1}_R=\int \mathrm{d}R_\alpha|R_\alpha\rangle\langle R_\alpha|$ and $\mathds{1}_P=\int \mathrm{d}P_\alpha|P_\alpha\rangle\langle P_\alpha|$, where $\alpha$ is the label of the imaginary-time (bead) index, and using the standard path-integral techniques,\cite{feynman1965,berne1986,ceperley1995} we obtain
\begin{align}\label{part}
{\cal{Z}} = &\frac{1}{(2\pi\hbar)^N} \lim_{N\rightarrow\infty} \int \mathrm{d}\{R_\alpha\} \int \mathrm{d}\{P_\alpha\}e^{-\beta_N{\tilde{H}_0}({\bf R})} \nonumber \\
&\times \mathrm{Tr_e}\Big[\prod_{\alpha=1}^Ne^{-\beta_N\frac{1}{\hbar}\mathbf{H}_\alpha\cdot\hat{\mathbf{S}}}\Big].
\end{align}
Here, we use the notation $\int \mathrm{d}\{X_\alpha\} = \prod_{\alpha=1}^N\int \mathrm{d}X_1\cdots \mathrm{d}X_N$, ${\bf R}\equiv\{R_\alpha\}$, and $\mathbf{H}_\alpha=[H_{x}(R_{\alpha}),H_{y}(R_{\alpha}),H_{z}(R_{\alpha})]$ (see their definition in Eq.~\ref{Hx}-\ref{Hz}). The state-independent ring polymer Hamiltonian $\tilde{H}_0$ is expressed as
\begin{align}\label{Hrp}
{\tilde{H}_0}({\bf R}) =& \sum_{\alpha=1}^N \big[\frac{P_\alpha^2}{2m} + \frac{m}{2\beta_N^2\hbar^2}(R_\alpha - R_{\alpha-1})^2 \\
&+ U_0(R_\alpha)+ \frac{1}{2}(V_1(R_\alpha) + V_2(R_\alpha))\big]. \nonumber
\end{align}
To perform the electronic trace, we insert $N$ copies of the following spin coherent state identities (by choosing $\mathrm{s=Q}$)
\begin{subequations}
\begin{align}
\mathds{1}_\mathbf{u} =& \int \mathrm{d}\mathbf{u_\alpha}|\mathbf{u_\alpha}\rangle\langle\mathbf{u_\alpha}|=\int \mathrm{d}\mathbf{u_\alpha}\hat{w}_\mathrm{Q} \\
=&\frac{1}{2\pi}\int_0^\pi \mathrm{d}\theta_\alpha\sin\theta_\alpha\int_0^{2\pi} \mathrm{d}\varphi_\alpha|\mathbf{u_\alpha}\rangle\langle\mathbf{u_\alpha}|,
\end{align}
\end{subequations}
and rearranging the terms (as well as neglecting a normalization constant), resulting in 
\begin{align}
{\cal{Z}} \propto& \lim_{N\rightarrow\infty} \int \mathrm{d}\{R_\alpha\} \int \mathrm{d}\{P_\alpha\} \int \mathrm{d} \{ \mathbf{u}_\alpha \}e^{-\beta_N\tilde{H}_0({\bf R})} \nonumber\\
&\times \prod_{\alpha=1}^N \langle\mathbf{u}_\alpha| e^{-\beta_N\frac{1}{\hbar}\mathbf{H}_\alpha\cdot\hat{\mathbf{S}}}|\mathbf{u}_{\alpha+1}\rangle.
\end{align}
The above partition function can also be equivalently expressed by inserting electronic projection operators $\hat{\mathcal P}=\sum_n|n\rangle\langle n|$, leading to
\begin{align}
{\cal{Z}} \propto &\lim_{N\rightarrow\infty} \int \mathrm{d}\{R_\alpha\} \int \mathrm{d}\{P_\alpha\} \int \mathrm{d} \{ \mathbf{u}_\alpha \} e^{-\beta_N\tilde{H}_0(\mathbf{R})} \nonumber\\
&\times \prod_{\alpha=1}^N \langle\mathbf{u}_\alpha|\sum_n|n\rangle\langle n| e^{-\beta_N\frac{1}{\hbar}\mathbf{H}_\alpha\cdot\hat{\mathbf{S}}}\sum_m|m\rangle\langle m|\mathbf{u}_{\alpha+1}\rangle.
\end{align}
Note that the size of the spin mapping Hilbert space $\mathds{1}_\mathbf{u}$ is the same as the original electronic subspace $\hat{\mathcal P}=\sum_n|n\rangle\langle n|$. Hence with or without $\hat{\mathcal{P}}$, the partition function is invariant. This is different than the mapping in harmonic oscillators based on the MMST formalism, where the mapping Hilbert space is larger than the original electronic subspace, and projection often leads to a better result.\cite{ananth2010,kelly2012}

We further express the matrix elements of spin coherent state projected by $\hat{\mathcal P}$ as follows
\begin{subequations}
\begin{align}
\mathbf{C}(\mathbf{u}_{\alpha})&\equiv\langle\mathbf{u}_\alpha|\sum_n|n\rangle\langle n| \\
&= \cos\frac{\theta_\alpha}{2}e^{i\frac{\varphi_\alpha}{2}}\langle 1| + \sin\frac{\theta_\alpha}{2}e^{-i\frac{\varphi_\alpha}{2}}\langle 2|, \nonumber\\
\mathbf{D}(\mathbf{u}_{\alpha+1})&\equiv\sum_m|m\rangle\langle m|\mathbf{u}_{\alpha+1}\rangle\\
&=\cos\frac{\theta_{\alpha+1}}{2}e^{-i\frac{\varphi_{\alpha+1}}{2}}|1\rangle + \sin\frac{\theta_{\alpha+1}}{2}e^{i\frac{\varphi_{\alpha+1}}{2}}|2\rangle. \nonumber
\end{align}
\end{subequations}
Using these, we can write the special form of the Spin Coherent State (SCS) partition function (with $\mathrm{s}=\mathrm{Q}$ case) as follows
\begin{equation}\label{Q-SCS}
{\cal{Z}} \propto \lim_{N\rightarrow\infty} \int \mathrm{d}\{R_\alpha\} \int \mathrm{d}\{P_\alpha\} \int \mathrm{d} \{\mathbf{u}_\alpha \}\mathrm{ Tr_{e}}[{\boldsymbol \Gamma}_\mathrm{Q}]\cdot e^{-\beta_N\tilde{H}_{0}({\bf R})},
\end{equation}
where the electronic trace has the following expression
\begin{subequations}
\begin{align}
&{\boldsymbol\Gamma}_\mathrm{Q} = \prod_{\alpha=1}^N\sum_{n,m}C_{n}(\mathbf{u}_{\alpha}){\cal{M}}_{nm}(R_{\alpha})D_{m}(\mathbf{u}_{\alpha+1}), \label{gammaQ}\\
&{\cal{M}}_{nm}(R_{\alpha}) = \langle n|e^{-\beta_N\frac{1}{\hbar}\mathbf{H}_\alpha\cdot\hat{\mathbf{S}}}|m\rangle.\label{M-mat}
\end{align}
\end{subequations}
This partition function is analogous to those used with MMST mapping variables, such as the mapping-variable RPMD partition function\cite{ananth2013} or the coherent state mapping (CSM) ring polymer partition function.\cite{chowdhury2017} In CSM partition function, a similar derivation procedure is conducted with the coherent-state representation of the MMST mapping oscillators.\cite{chowdhury2017,hsieh2012}

The above procedure relies on inserting N copies of the identities $\mathds{1}_\mathbf{u}=\int \mathrm{d}\mathbf{u_\alpha}|\mathbf{u_\alpha}\rangle\langle\mathbf{u_\alpha}| \equiv \int \mathrm{d}\mathbf{u_\alpha}\hat{w}_\mathrm{Q}(\mathbf{u}_\alpha)$ (where $\alpha$ is the bead index). Of course, one can insert the general resolution of identity $\mathds{1}_\mathbf{u} = \int \mathrm{d}\mathbf{u}\hat{w}_{\bar{\mathrm{s}}}$ (Eq.~\ref{resolution}) inside the  $\mathrm{Tr_e}[...]$ of Eq.~\ref{part}, then moving the $\int \mathrm{d} \mathbf{u}_\alpha$ integral outside $\mathrm{Tr_e}$, resulting in 
\begin{equation}\label{SCS-general}
{\cal{Z}} \propto \lim_{N\rightarrow\infty} \int \mathrm{d}\{R_\alpha\} \int \mathrm{d}\{P_\alpha\} \int \mathrm{d} \{ \mathbf{u}_\alpha \}\mathrm{Tr_{e}}[{\boldsymbol\Gamma}_{\mathrm{s}}]\cdot e^{-\beta_N\tilde{H}_0({\bf R})},
\end{equation}
where the expression of the electronic trace is 
\begin{equation}\label{gammas}
{\bf \Gamma}_{\mathrm{s}} =\prod_{\alpha=1}^N  e^{-\beta_N\frac{1}{\hbar}\mathbf{H}_\alpha\cdot\hat{\mathbf{S}}}\cdot\hat{w}_{\mathrm{s}}(\mathbf{u}_\alpha).
\end{equation}
By Taylor expanding the Boltzmann operator and using the properties of the Pauli matrices, we can prove the following identity
\begin{equation}\label{eq:Taylorexp}
e^{-\beta_N\frac{1}{\hbar}\mathbf{H}_\alpha\cdot\hat{\mathbf{S}}} = \cosh\frac{\beta_N|\mathbf{H}_\alpha|}{2}\hat{\mathcal{I}}-\sinh\frac{\beta_N|\mathbf{H}_\alpha|}{2}\cdot\frac{2\mathbf{H}_\alpha\cdot\hat{\mathbf{S}}}{\hbar|\mathbf{H}_\alpha|},
\end{equation}
where $|\mathbf{H}_\alpha|=\sqrt{H_x^2(R_\alpha)+H_y^2(R_\alpha)+H_z^2(R_\alpha)}$. Plugging this identity back into Eq.~\ref{gammas} we obtain the general expression ${\boldsymbol\Gamma}_{\mathrm{s}}$ as follows
\begin{align}\label{eq:SCSgen}
\boldsymbol{\Gamma}_\mathrm{s}=&\Big[\prod_{\alpha=1}^N\Big(\frac{1}{2}\cosh\frac{\beta_N|\mathbf{H}_\alpha|}{2}-r_{\mathrm{s}}\frac{\mathbf{H}_\alpha}{|\mathbf{H}_\alpha|}\cdot\mathbf{u}_\alpha\sinh\frac{\beta_N|\mathbf{H}_\alpha|}{2}\Big)\hat{\mathcal{I}}\nonumber\\
&+\Big(r_\mathrm{s}\mathbf{u}_\alpha\cosh\frac{\beta_N|\mathbf{H}_\alpha|}{2}\\
&~~~~~-\frac{1}{|\mathbf{H}_\alpha|}(\frac{\mathbf{H}_\alpha}{2}+ir_{\mathrm{s}}\mathbf{H}_\alpha\wedge\mathbf{u}_\alpha)\sinh\frac{\beta_N|\mathbf{H}_\alpha|}{2}\Big)\cdot\hat{\boldsymbol{\sigma}}\Big].\nonumber
\end{align}
A detailed derivation of Eqs.~\ref{eq:Taylorexp} and~\ref{eq:SCSgen} is provided in Appendix D. When $\mathrm{s=Q}$, Eq.~\ref{eq:SCSgen} is equivalent to the expression of ${\boldsymbol\Gamma}_{\mathrm{Q}}$ in Eq.~\ref{Q-SCS}. The numerical advantage of Eq.~\ref{eq:SCSgen} is that it replaces the $\mathcal{M}_{nm}(R_{\alpha})$ matrix in Eq.~\ref{M-mat} with an analytic expression in Eq.~\ref{eq:Taylorexp}.

\subsection{Spin-Mapping (SM)-NRPMD Hamiltonian}
The SCS partition function in Eq.~\ref{SCS-general} gives the exact quantum statistics for a non-adiabatic system. The effective Hamiltonian from the SCS partition function can be used to propagate the dynamics. However, it will not provide accurate electronic dynamics (such as electronic Rabi oscillation) due to the inter-bead coupling among the different electronic and nuclear DOFs inside ${\boldsymbol\Gamma}_\mathrm{s}$.

Instead of proposing a reasonable Hamiltonian for dynamics propagation, here, we try to theoretically justify a Hamiltonian from an alternative expression of the partition function. To this end, we evaluate the electronic trace in Eq.~\ref{part} using the property  $\mathrm{Tr_e}[\hat{A}]=\int \mathrm{d}\mathbf{u}A_\mathrm{s}(\mathbf{u})$ in Eq.~\ref{trA}, leading to 
\begin{align}
\mathrm{Tr_e}\Big[\prod_{\alpha=1}^Ne^{-\beta_N\frac{1}{\hbar}\mathbf{H}_\alpha\cdot\hat{\mathbf{S}}}\Big]= \int \mathrm{d}\mathbf{u}_1 \Big[\prod_{\alpha=1}^Ne^{-\beta_N\frac{1}{\hbar}\mathbf{H}_\alpha\cdot\hat{\mathbf{S}}}\Big]_{\mathrm{s}}(\mathbf{u}_1), 
\end{align}
where $\mathrm{s}\in\{\mathrm{Q,P,W}\}$. We further separate $\prod_{\alpha=1}^Ne^{-\beta_N\frac{1}{\hbar}\mathbf{H}_\alpha\cdot\hat{\mathbf{S}}}$ into $e^{-\beta_N\frac{1}{\hbar}\mathbf{H}_1\cdot\hat{\mathbf{S}}}\prod_{\alpha=2}^Ne^{-\beta_N\frac{1}{\hbar}\mathbf{H}_\alpha\cdot\hat{\mathbf{S}}}$, and use the property expressed in Eq.~\ref{eq:QPQ}, leading to 
\begin{align}\label{eq:SWS}
&\int \mathrm{d}\mathbf{u}_1 \Big[e^{-\beta_N\frac{1}{\hbar}\mathbf{H}_1\cdot\hat{\mathbf{S}}}\prod_{\alpha=2}^Ne^{-\beta_N\frac{1}{\hbar}\mathbf{H}_\alpha\cdot\hat{\mathbf{S}}}\Big]_{\mathrm{s}}(\mathbf{u}_1)\\
&=\int \mathrm{d}\mathbf{u}_1 \Big[e^{-\beta_N\frac{1}{\hbar}\mathbf{H}_1\cdot\hat{\mathbf{S}}}\Big]_{\mathrm{s}}(\mathbf{u}_1)\cdot\Big[\prod_{\alpha=2}^Ne^{-\beta_N\frac{1}{\hbar}\mathbf{H}_\alpha\cdot\hat{\mathbf{S}}}\Big]_{\bar{\mathrm{s}}}(\mathbf{u}_1), \nonumber
\end{align}
where $\{\mathrm{s},\bar{\mathrm{s}}\}$ can be any pair that is permitted based on Eq.~\ref{eq:QPQ}.

To evaluate $[e^{-\beta_N\frac{1}{\hbar}\mathbf{H}_1\cdot\hat{\mathbf{S}}}]_{\mathrm{s}}(\mathbf{u}_1)$, we Taylor expand the exponential and neglect the terms of order equals to or higher than $\beta_N^2$ (which is exact under the limit $N\rightarrow\infty$), leading to 
\begin{align}\label{boltz-lin}
&[1-\beta_N \frac{1}{\hbar}\mathbf{H}_1\cdot\hat{\mathbf{S}} + {\cal{O}}(\beta_N^2)]_{\mathrm{s}}(\mathbf{u}_1)\\
&= \exp[-\beta_N \cdot\frac{1}{\hbar}\mathbf{H}_1\cdot[\hat{\mathbf{S}}]_{\mathrm{s}}(\mathbf{u}_1)]=\exp[-\beta_N\cdot\frac{1}{\hbar} r_\mathrm{s}\mathbf{H}_1\cdot\mathbf{u}_1]\nonumber.
\label{eq:swstaylor}
\end{align}
Plugging it back into Eq.~\ref{eq:SWS}, we have
\begin{align}
&\mathrm{Tr_e}\Big[\prod_{\alpha=1}^Ne^{-\beta_N\frac{1}{\hbar}\mathbf{H}_\alpha\cdot\hat{\mathbf{S}}}\Big]\\
&=\int \mathrm{d}\mathbf{u}_1 e^{-\beta_N\cdot r_\mathrm{s}\mathbf{H}_1\cdot\mathbf{u}_1}\cdot \Big[\prod_{\alpha=2}^Ne^{-\beta_N\frac{1}{\hbar}\mathbf{H}_\alpha\cdot\hat{\mathbf{S}}}\Big]_{\bar{\mathrm{s}}}(\mathbf{u}_1) \nonumber\\
&=\int \mathrm{d}\mathbf{u}_1 e^{-\beta_N\cdot r_\mathrm{s}\mathbf{H}_1\cdot\mathbf{u}_1}\cdot\mathrm{Tr_{e}}\Big[\prod_{\alpha=2}^Ne^{-\beta_N\frac{1}{\hbar}\mathbf{H}_\alpha\cdot\hat{\mathbf{S}}}\hat{w}_{\bar{\mathrm{s}}}(\mathbf{u}_1)\Big].\nonumber
\label{eq:PtoQ1}
\end{align}
Further inserting the identity $\int \mathrm{d} \mathbf{u}_2\hat{w}_\mathrm{s}(\mathbf{u}_2)$ (see Eq.~\ref{resolution}) inside the $\mathrm{Tr_{e}}$, we have 
\begin{align}
&\mathrm{Tr_e}\Big[\prod_{\alpha=1}^Ne^{-\beta_N\frac{1}{\hbar}\mathbf{H}_\alpha\cdot\hat{\mathbf{S}}}\Big]\\
&=\int \mathrm{d}\mathbf{u}_1 e^{-\beta_N\cdot r_\mathrm{s}\mathbf{H}_1\cdot\mathbf{u}_1}\cdot\mathrm{Tr_{e}}\Big[\int \mathrm{d} \mathbf{u}_2\hat{w}_\mathrm{s}(\mathbf{u}_2)e^{-\beta_N\frac{1}{\hbar}\mathbf{H}_2\cdot\hat{\mathbf{S}}}\nonumber\\
&~~~~~~~~~~~~~~~~~~~~~~~~~~~~~~~~~~~~\times\prod_{\alpha=3}^Ne^{-\beta_N\frac{1}{\hbar}\mathbf{H}_\alpha\cdot\hat{\mathbf{S}}}\hat{w}_{\bar{\mathrm{s}}}(\mathbf{u}_1)\Big]\nonumber\\
&=\int \mathrm{d}\mathbf{u}_1 e^{-\beta_N\cdot r_\mathrm{s}\mathbf{H}_1\cdot\mathbf{u}_1}\nonumber\\
&~~~~\times\int \mathrm{d} \mathbf{u}_2\Big[e^{-\beta_N\frac{1}{\hbar}\mathbf{H}_2\cdot\hat{\mathbf{S}}}\prod_{\alpha=3}^Ne^{-\beta_N\frac{1}{\hbar}\mathbf{H}_\alpha\cdot\hat{\mathbf{S}}}\hat{w}_{\bar{\mathrm{s}}}(\mathbf{u}_1)\Big]_\mathrm{s}(\mathbf{u}_2)\nonumber\\
&=\int \mathrm{d}\mathbf{u}_1 e^{-\beta_N\cdot r_\mathrm{s}\mathbf{H}_1\cdot\mathbf{u}_1}\int \mathrm{d} \mathbf{u}_2e^{-\beta_N\cdot r_\mathrm{s}\mathbf{H}_2\cdot\mathbf{u}_2} \nonumber\\
&~~~~\times\mathrm{Tr_e}\Big[\prod_{\alpha=3}^Ne^{-\beta_N\frac{1}{\hbar}\mathbf{H}_\alpha\cdot\hat{\mathbf{S}}}\hat{w}_{\bar{\mathrm{s}}}(\mathbf{u}_1)\hat{w}_{\bar{\mathrm{s}}}(\mathbf{u}_2)\Big],\nonumber
\label{eq:PtoQ2}
\end{align}
where in the second equality, we have moved the $\int \mathrm{d} \mathbf{u}_2$ outside the $\mathrm{Tr_{e}}$, as well as used the definition of $[\hat{A}]_\mathrm{s}(\mathbf{u})$ in Eq.~\ref{eq:sfuncdef}, and in the third equality, we have used the property in Eq.~\ref{eq:SWS}.

Repeating the above argument for all $N$ beads, we obtain the following partition function
\begin{equation}
{\cal{Z}} \propto \lim_{N\rightarrow\infty} \int \mathrm{d}\{R_\alpha\} \int \mathrm{d}\{P_\alpha\}\int \mathrm{d} \{ \mathbf{u}_\alpha \}\Phi_{\bar{\mathrm{s}}}\cdot e^{-\beta_N \tilde{H}_\mathrm{s}} ,
\end{equation}
where $\Phi_{\bar{\mathrm{s}}}=\mathrm{Tr_e}\left[\prod_{\alpha=1}^N\hat{w}_{\bar{s}}(\mathbf{u}_\alpha)\right]$, and the spin-mapping (SM)-NRPMD Hamiltonian is 
\begin{equation}\label{SW-NRP}
\tilde{H}_\mathrm{s} = \tilde{H}_0({\bf R}) + \sum_{\alpha=1}^N r_\mathrm{s} \mathbf{H}_\alpha\cdot\mathbf{u}_\alpha,
\end{equation}
which is the ring polymer generalization of $H_\mathrm{s}(\mathbf{u})$ in Eq.~\ref{eq:sm-ham} (with the additional ring polymer potential in Eq.~\ref{Hrp}). Based on our previous experience with the MMST version of NRPMD approach, we conjecture that $\tilde{H}_\mathrm{s}$ should be the Hamiltonian for the NRPMD propagation when using the spin mapping variables. This is because the correct equations of motion for the MMST mapping variables\cite{hele2016,chowdhury2021} can be derived based on the partition function\cite{chowdhury2019} through a similar procedure as above, which coincides with the Liouvillian derived from generalized Kubo-transformed TCF with Matsubara approximation and ring polymer approximation\cite{chowdhury2021}  We also note that in principle, the partition function in Eq.~\ref{SW-NRP} should generate the same result as the one in Eq.~\ref{SCS-general}, under the limit $N\to\infty$. However, with a finite $N$, we find that the numerical convergence by using Eq.~\ref{SW-NRP} is much slower compared to Eq.~\ref{SCS-general}, likely due to the limit we took in Eq.~\ref{boltz-lin} (which requires a large $N$). Hence, we emphasize that Eq.~\ref{SW-NRP} is only used as a justification for the Stratonovich-Weyl NRPMD Hamiltonian in Eq.~\ref{SW-NRP}, and not used for sampling the quantum initial condition.

\section{Spin-Mapping (SM)-NRPMD Time-Correlation Function}
The Kubo-transform real-time correlation function for two operators $\hat{A}$ and $\hat{B}$ is expressed as 
\begin{align}\label{kubo}
C^\mathrm{K}_{AB}(t) = \frac{1}{\cal{Z}\beta}\int_0^\beta \mathrm{d}\lambda\mathrm{Tr}\left[e^{-(\beta-\lambda)\hat{H}}\hat{A}e^{-\lambda\hat{H}}e^{i\hat{H}t/\hbar}\hat{B}e^{-i\hat{H}t/\hbar}\right].
\end{align}

We propose that the above Kubo-transformed TCF (Eq.\ref{kubo}) can be approximated as the following Spin-Mapping TCF
\begin{align}
C_{AB}(t) = &\frac{1}{\cal{Z}}\lim_{N\rightarrow\infty} \int \mathrm{d}\{R_\alpha\} \int \mathrm{d}\{P_\alpha\} \int \mathrm{d}\{ \mathbf{u}_\alpha \} \nonumber\\
&~~~\times\mathrm{Tr_{e}}[\boldsymbol{\Gamma}_\mathrm{s}]e^{-\beta_N\tilde{H}_{0}}[A]_{N}(0)[B]_{N}(t),
\end{align}
where $[A]_{N}(0)=\frac{1}{N}\sum_{\alpha=1}^{N}A(R_{\alpha})\equiv \bar{A}(0)$ and $[B]_{N}(t)=\frac{1}{N}\sum_{\alpha=1}^{N}B(R_{\alpha}(t))\equiv\bar{B}(t)$ for $\hat{A}(\hat{R})$ and $\hat{B} (\hat{R})$ when they are functions of $\hat{R}$. When operators $\hat{A}$ and $\hat{B}$ are related to the electronic DOF, the TCF is proposed as
\begin{align}
C_{AB}(t) = &\frac{1}{\cal{Z}}\lim_{N\rightarrow\infty} \int \mathrm{d}\{R_\alpha\} \int \mathrm{d}\{P_\alpha\} \int \mathrm{d}\{ \mathbf{u}_\alpha \} \nonumber\\
& \times \mathrm{Tr_{e}}[\boldsymbol{\Gamma}_\mathrm{s}\hat{A}]e^{-\beta_N\tilde{H}_{0}}[B_{\bar{\mathrm s}}]_{N}(t),
\end{align}
with $\{\mathrm{s},\bar{\mathrm s}\}$ being complementary indexes permitted by Eq.~\ref{eq:QPQ} in order to satisfy the requirement at $t=0$ to compute the trace of two operators (i.e, $e^{-\beta H}\hat{A}$ and $\hat{B}$). The population estimator ${\cal{P}}^\mathrm{s}_{nn}$ for the operator $\hat{A} = |n\rangle\langle n|$ is obtain with $\mathrm{Tr_{e}}[{\boldsymbol\Gamma}_\mathrm{s}|n\rangle\langle n|]$, which one can write in a bead-averaged fashion
\begin{align}\label{pnn}
&{\mathcal P}^\mathrm{s}_{nn}=\mathrm{Tr_{e}}[{\boldsymbol\Gamma}_\mathrm{s}|n\rangle\langle n|]=\frac{1}{N}\sum_{\mu=1}^{N}\Big[\prod_{\alpha'=1}^{N-\mu}  e^{-\beta_N\frac{1}{\hbar}\mathbf{H}_{\alpha'}\cdot\hat{\mathbf{S}}}\cdot\hat{w}_{\mathrm{s}}(\mathbf{u}_{\alpha'})\nonumber\\
&~~~~~~\times|n\rangle\langle n|\prod_{\alpha''=N-\mu+1}^{N}  e^{-\beta_N\frac{1}{\hbar}\mathbf{H}_{\alpha''}\cdot\hat{\mathbf{S}}}\cdot\hat{w}_{\mathrm{s}}(\mathbf{u}_{\alpha''})\Big], 
\end{align}
to improve the statistical convergence. The analytic expression can be evaluated in the same way as ${\boldsymbol\Gamma}_\mathrm{s}$ in Eq.~\ref{eq:SCSgen}, leading to $|n\rangle\langle n|$ inserted in-between the $\alpha$ and the $\alpha+1$ bead. Specifically, for $\mathrm{s=Q}$, using Eq.~\ref{gammaQ} we have
\begin{align}
{\cal P}^\mathrm{Q}_{nn} = &\frac{1}{N}\sum_{\alpha=1}^N\frac{\sum_{m}C_{n}({\bf u}_{\alpha}){\cal{M}}_{nm}(R_{\alpha})D_{m}({\bf u}_{\alpha+1})}{\sum_{n,m}C_{n}({\bf u}_{\alpha}){\cal{M}}_{nm}(R_{\alpha})D_{m}({\bf u}_{\alpha+1})}.
\label{eq:scsest}
\end{align}

The population estimator for the operator $\hat{B}$ is obtained by 
\begin{equation}
[B_{\bar{\mathrm s}}]_{N}=\frac{1}{N}\sum_{\alpha=1}^{N}B_{\bar{\mathrm{s}}}({\bf u}_{\alpha}),
\end{equation}
where $B_{\bar{\mathrm{s}}}({\bf u}_{\alpha})$ is the SW transform of $\hat{B}$, and when $\hat{B}=|n\rangle\langle n|$, it is expressed as
\begin{align}
B_{\bar{\mathrm s}} ({\bf u}_{\alpha}) = \mathrm{Tr_e}\left[|n\rangle\langle n|\hat{w}_{\bar{s}}\right] = \begin{cases}
1/2 + r_{\bar{\mathrm s}}\cos\theta_\alpha, & n=1 \\ 
1/2 - r_{\bar{\mathrm s}}\cos\theta_\alpha, & n=2
\end{cases}.\nonumber
\end{align}
The function $[B]_{N}(t)$ or $[B_{\bar{\mathrm{s}}}]_{N}(t)$ is evaluated along the classical trajectory $\{R_{\alpha}(t),{\bf u}_{\alpha}(t)\}$, and the dynamics is {\it proposed} to be governed by 
\begin{equation}
\tilde{H}_{\bar{\mathrm s}} = \tilde{H}_0({\bf R}) + \sum_{\alpha=1}^N r_{\bar{\mathrm{s}}} \mathbf{H}(R_\alpha)\cdot\mathbf{u}_\alpha
\end{equation}
where the Hamiltonian is justified in Eq.~\ref{SW-NRP}. The equations of motion are expressed as
\begin{subequations}
\begin{align}
\dot{R}_\alpha &= \frac{\partial \tilde{H}_{\bar{\mathrm{s}}}}{\partial P_\alpha} = \frac{P_\alpha}{m}\label{EOM-SM1}\\
\dot{P}_\alpha &= - \frac{\partial \tilde{H}_{\bar{\mathrm{s}}}}{\partial R_\alpha} = - \frac{\partial \tilde{H}_0}{\partial R_\alpha} - r_{\bar{\mathrm{s}}}\frac{\partial {\bf H}(R_{\alpha})}{\partial R_\alpha}\cdot\mathbf{u}_\alpha,\label{EOM-SM2}\\
\dot{{\bf u}}_{\alpha}&=\frac{1}{\hbar}{\bf H}(R_{\alpha})\wedge{\bf u}_{\alpha}.\label{EOM-SM3}
\end{align}
\end{subequations}
In the original NRPMD method, the corresponding equation of motion was first proposed,\cite{richardson2013} then recently proved through the non-adiabatic Matsubara dynamics formalism.\cite{chowdhury2021} We envision that the above EOM (Eq.~\ref{EOM-SM1}-Eq.~\ref{EOM-SM3}) can also be proved in a similar way when using the spin mapping variables, and we will explore this in future studies.

\section{Computational Details}
To test the performance of the derived SCS-partition function in Eq.~\ref{SCS-general}, we adapt a widely used model system\cite{Alexander2001,tully2007, ananth2010} and compute the state-dependent nuclear probability distribution. The model Hamiltonian $\hat{H}=\hat{P}^2/2m+\hat{V}$, with nuclear mass $M=3600$ a.u., and the diabatic potential $\hat{V}$ is defined as
\begin{align}\label{R-model}
V_{ij} = \begin{cases}
\frac{1}{2}k_i(R-R_i)^2+\epsilon_i, & i=j \\ 
5\times10^{-5}e^{-0.4R^2}, & i\neq j
\end{cases},
\end{align}
where the model parameters are presented in Table.~\ref{tab:nucd}. We refer to this model as Model 0. The physical temperature of the system is set to be $T=8$ K.

The initial quantum distribution is sampled using the Metropolis-Hastings algorithm according to the following distribution function
\begin{equation}\label{rho}
\rho(\{R_{\alpha}, {\bf u}_{\alpha}\})=|\mathrm{Tr_{e}}[{\boldsymbol\Gamma}_{\mathrm{s}}]|\cdot e^{-\beta_N\tilde{H}_0({\bf R})},
\end{equation}
with a complex weighting factor of
\begin{equation}
\Xi_\mathrm{s}(\{R_{\alpha}, {\bf u}_{\alpha}\})=\mathrm{Tr_{e}}[{\boldsymbol\Gamma}_{\mathrm{s}}]/|\mathrm{Tr_{e}}[{\boldsymbol\Gamma}_{\mathrm{s}}]|.
\end{equation}

The nuclear probability distribution is obtained by computing
\begin{align}
\mathrm{P}(R_{0}) =& \frac{\mathrm{Tr}[e^{-\beta \hat{H}}\delta(\hat{R}-R_0)]}{\mathcal Z} \\
=&\frac{1}{\langle\Re(\Xi_\mathrm{s})\rangle}\cdot\langle \Re(\Xi_\mathrm{s})\cdot\delta(R-R_{0})\rangle,\nonumber
\end{align}
where $\mathrm{Tr}=\mathrm{Tr_{n}Tr_{e}}$ (trace over both nuclear and electronic DOFs), and the bracket $\langle... \rangle$ indicates the ensemble average with respect to $\rho(\{R_{\alpha}, {\bf u}_{\alpha}\})$ in Eq.~\ref{rho}.
The state-resolved probability distribution is obtained by projecting the distribution onto a given state $|n\rangle\langle n|$ leading to the probability
\begin{align}
\mathrm{P}_{n}(R_0)&=\frac{\mathrm{Tr}[e^{-\beta \hat{H}}|n\rangle\langle n|\delta(\hat{R}-R_0)]}{\mathcal Z}\\ &=\frac{1}{\langle\Re(\Xi_\mathrm{s})\rangle}\cdot\langle\Re(\Xi_\mathrm{s}\cdot{\mathcal P}^\mathrm{s}_{nn})\cdot\delta(R-R_{0})\rangle,\nonumber
\end{align}
with the estimator ${\cal{P}}^\mathrm{s}_{nn}$ expressed in Eq.~\ref{pnn}. To compute $\mathrm{P}(R_0)$ and $\mathrm{P}_{n}(R_0)$, $N=10$ beads were required to converge the results, using a total of $2.4\times10^7$ configurations sampled from the Monte-Carlo procedure for $\mathrm{s}=\mathrm{Q}$. Exactly identical results can be obtain with the same bead-convergence for other choices of s, but the required number of configurations to achieve the same level of convergence is much higher. In particular for $N=10$ beads, using $\mathrm{s}=\mathrm{W}$ requires 24 times more trajectories, while using $\mathrm{s}=\mathrm{P}$ requires almost 2000 times more trajectories.

\begin{table}[htp]
\caption{Parameters for Model 0}
\begin{tabular*}{\columnwidth}{c @{\extracolsep{\fill}} c @{\extracolsep{\fill}} c}
\hline\hline
$i$ & 1 & 2 \\
\hline
$k_i$ & \hspace{0.2cm}$4\times10^{-5}$\hspace{0.2cm} & \hspace{0.2cm}$3.2\times10^{-5}$\hspace{0.2cm} \\
$R_i$ & $-1.75$ & $1.75$ \\
$\epsilon_i$ & $0.0$ & $2.28\times10{-5}$ \\ [0.5ex]
\hline\hline
\end{tabular*}
\label{tab:nucd}
\end{table}

To assess the accuracy of the SM-NRPMD approach, we compute time correlation functions and compare our results with numerically exact Kubo-transformed quantum TCF, as well as non-adiabatic RPMD approach based on the MMST formalism.\cite{richardson2013} The model used for those calculations is a simple two-level system linearly coupled to a harmonic potential
\begin{equation}
\hat{H} = \frac{\hat{P}^2}{2m} + \frac{1}{2}m\omega^2\hat{R}^2 + \begin{pmatrix}
\hat{R} + \epsilon & \Delta \\ 
\Delta & -\hat{R} - \epsilon
\end{pmatrix},
\end{equation}
where $\Delta$ is a constant electronic coupling and $2\epsilon$ is the energy bias between the two electronic states. We choose $m=\hbar=\omega=\beta=1$. The rest of the parameters are provided in Table~\ref{tab:modelparam}, changing the non-adiabaticity of the system from adiabatic (model I with $\beta\Delta = 10$) to highly non-adiabatic (model VII with $\beta\Delta = 0.1$). The number of beads to generate the converged results are also provided in Table~\ref{tab:modelparam}.

The position and population auto-correlation functions are computed as follows
\begin{align}
\mathrm{C}_{RR}(t) =& \frac{1}{\langle\Re(\Xi_\mathrm{s})\rangle}\cdot {\langle\Re(\Xi_\mathrm{s})\cdot\bar{R}(0)\cdot\bar{R}(t)\rangle}, \\
\mathrm{C}_{nn}(t) =& \frac{1}{\langle\Re(\Xi_\mathrm{s}) \rangle}\langle\Re(\Xi_\mathrm{s}\cdot{\cal{P}}^\mathrm{s}_{nn}(0))\cdot[B_{\bar{\mathrm s}}]_{N}(t)\rangle.
\end{align}
For $\mathrm{s}=\mathrm{W}$, between $10^4$ and $10^6$ trajectories were run for 4-to-6 beads for the results presented hereafter, with a time-step of 0.01 a.u.

\begin{table}[htp]
\caption{Parameters for models I-VII}
\begin{tabular*}{\columnwidth}{c @{\extracolsep{\fill}} c @{\extracolsep{\fill}} c @{\extracolsep{\fill}} c @{\extracolsep{\fill}} c @{\extracolsep{\fill}} c @{\extracolsep{\fill}} c @{\extracolsep{\fill}} c @{\extracolsep{\fill}} c}
\hline\hline
Models & I & II & III & IV & V & VI & VII\\
\hline
$\Delta$ & 10 & 4 & 1 & 1 & 1 & 0.1 & 0.1\\
$\epsilon$ & \hspace{0.2cm} 0 \hspace{0.2cm} & \hspace{0.2cm} 0 \hspace{0.2cm} & \hspace{0.2cm} 2 \hspace{0.2cm} & \hspace{0.2cm} 0.5 \hspace{0.2cm} & \hspace{0.2cm} 0 \hspace{0.2cm} & \hspace{0.2cm} 1.5 \hspace{0.2cm} & \hspace{0.2cm} 0 \hspace{0.2cm}\\
beads & 4 & 6 & 6 & 6 & 6 & 6 & 4 \\ [0.5ex]
\hline\hline
\end{tabular*}
\label{tab:modelparam}
\end{table}
\begin{figure}[htp]
	\includegraphics[width=\columnwidth]{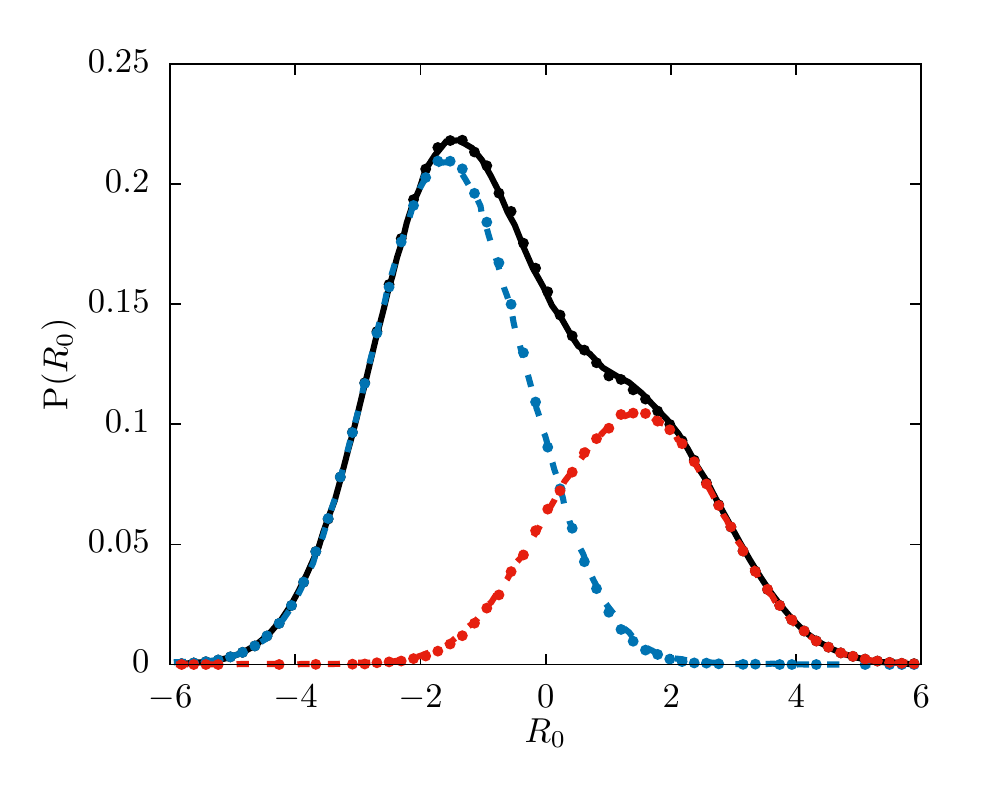}
    \caption {Nuclear probability distribution (black curve) $\mathrm{P}(R_{0})$  of model 0 obtained from the SCS partition function, with $N=10$ beads. The state specific distributions for state 1 (blue) and state 2 (red) are also shown. The results are compared with quantum exact calculations (filled circles).}
	\label{fig:nucd}
\end{figure}

\section{Results and discussion}  
Fig.~\ref{fig:nucd} presents the nuclear probability distribution P$(R_0)$ (black) as well as the state-resolved nuclear probability distributions P$_{1}(R_0)$ (blue) and P$_{2}(R_0)$ (red) for a widely used model system described in Eq.~\ref{R-model}. These distributions agree perfectly with the numerically exact results obtained from the DVR calculations\cite{colbert1992}. The numerical convergence is achieved with only $N=10$ beads. The SCS partition function in Eq.~\ref{SCS-general} only requires two independent variables $\{\theta_{\alpha}, \varphi_{\alpha}\}$ for each bead, which is consistent with the number of electronic states. The MMST based partition function, such as those used in NRPMD or MV-RPMD requires 4 independent variables. As the number of beads increases, the MMST-based approaches becomes numerically expensive. In addition, previous numerical investigations suggest that 16-32 beads are required to reach to the same level of convergence with the MMST-based path-integral approaches.\cite{ananth2010} This is likely due to a larger set of free variables needed to be sampled. Moreover, the general formalism of ${\boldsymbol\Gamma}_\mathrm{s}$ in Eq.~\ref{eq:SCSgen} does not explicitly require the evaluation of ${\cal{M}}_{nm}(R_{\alpha}) = \langle n|e^{-\beta_N\frac{1}{\hbar}\mathbf{H}_\alpha\cdot\hat{\mathbf{S}}}|m\rangle$ matrix, avoiding explicit diagonalization of the $2\times2$ matrix at a given $R_{\alpha}$. 
\begin{figure}[htp]
\includegraphics[width=\columnwidth]{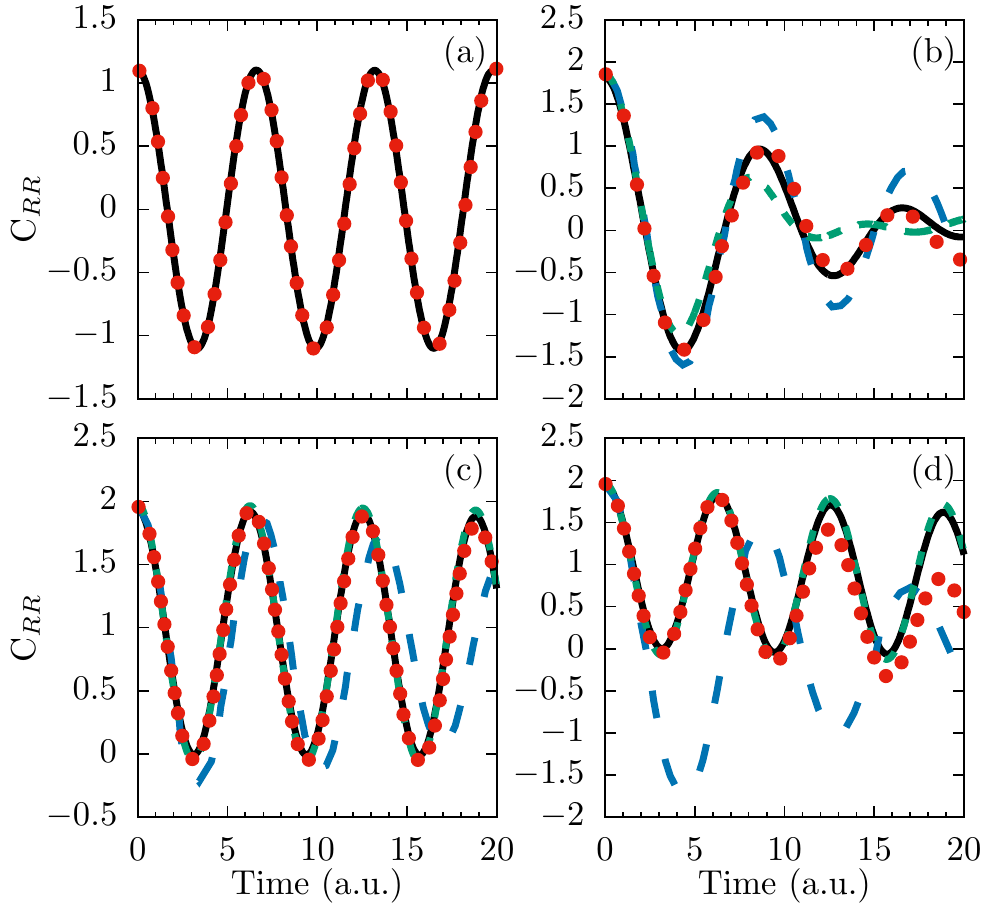}
\caption {Kubo-transformed nuclear position auto-correlation functions for (a) model I, (b) model V, (c) model VI, and (d) model VII. Results are obtained from SM-NRPMD with $\mathrm{s}=\mathrm{W}$ (black lines), MF-RPMD (blue dashed lines) and NRPMD\cite{richardson2013} (green dashed lines), as well as numerically exact result (red dots).}
	\label{fig:crr}
\end{figure}

\begin{figure}[htp]
\includegraphics[width=\columnwidth]{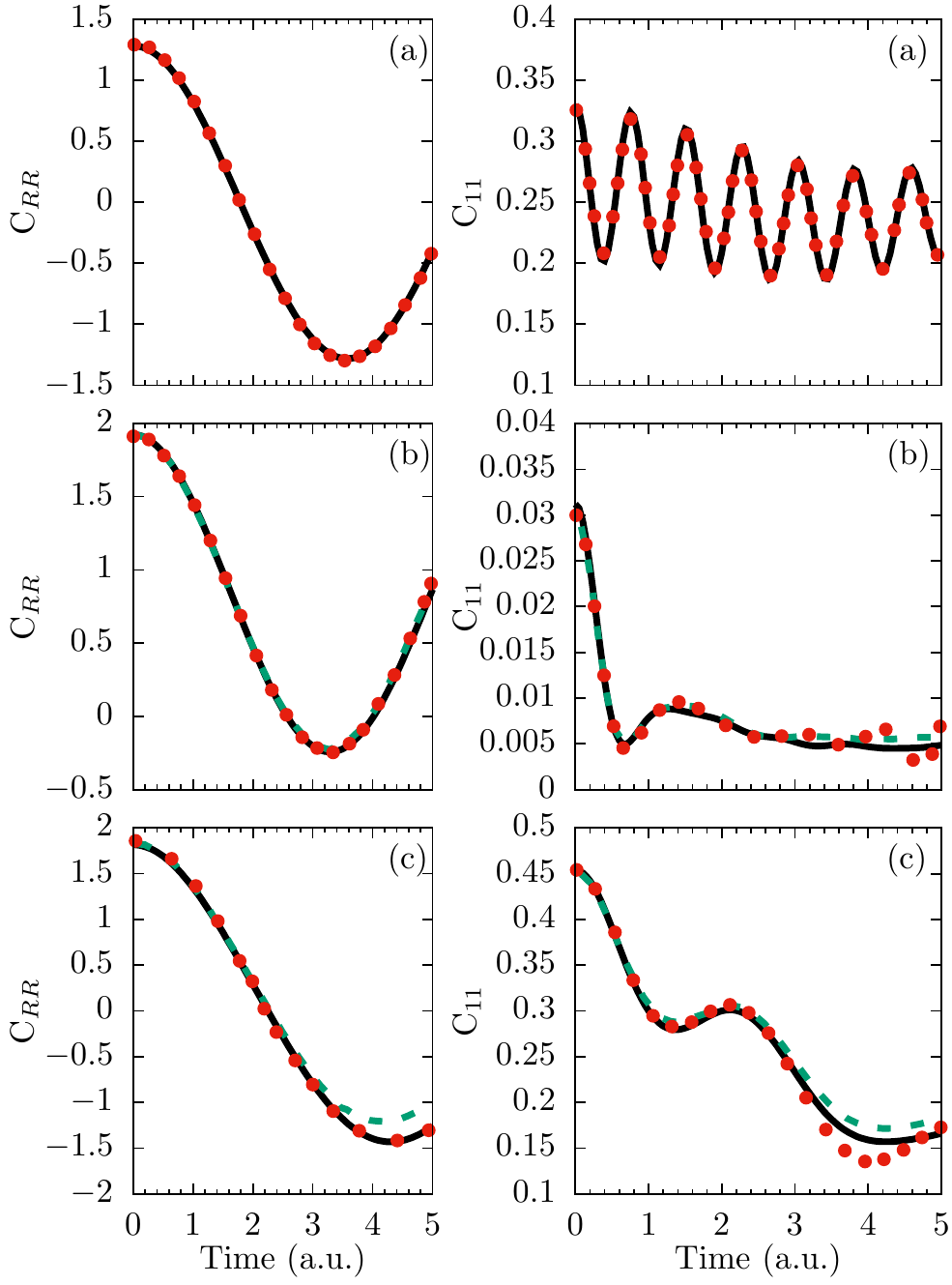}
\caption {The Kubo-transformed nuclear position auto-correlation functions (left panels) and the electronic population auto-correlation functions (right panels) for (a) Model II, (b) Model III, and (c) Model V. The results are obtained from SM-NRPMD with $\mathrm{s}=\mathrm{W}$ (black lines) and NRPMD\cite{richardson2013} (green dashed lines), compared to the numerically exact result (red dots).}
	\label{fig:cnn}
\end{figure}

Fig.~\ref{fig:crr} presents the nuclear position auto-correlation function computed from SM-NRPMD (black), NRPMD\cite{richardson2013,richardson2017} (blue dashed), mean-field RPMD\cite{ananth2013,hele2011} (blue),  and the numerically exact method (red dots). A brief description of NPRMD and the mean-field RPMD approach are provided in Appendix E. The SM-NRPMD calculations have been done with the choice of $\mathrm{s}=\bar{\mathrm{s}}=\mathrm{W}$ (sampling with $\boldsymbol{\Gamma}_\mathrm{s}\equiv\boldsymbol{\Gamma}_\mathrm{W}$ and dynamics with $\tilde{H}_{\bar{\mathrm{s}}}\equiv\tilde{H}_\mathrm{W}$). In the adiabatic regime ($\beta \Delta \gg 1$) in panel a, all methods agree perfectly with the exact result as expected.  In the intermediate regime $\beta\Delta\approx 1$ in panel b, all RPMD based approaches captures the correct oscillation frequency of the TCF, but they give different amplitudes that deviate from the exact results, except the SM-NRPMD approach which provides an excellent agreement with the exact results. In the non-adiabatic regime though ($\beta\Delta\gg1$) in panels c and d, MF-RPMD method can not provide the correct  amplitude nor oscillation frequency for the TCF. On the other hand, both SM-NRPMD and NRPMD results are in agreement with the exact results at short times. We can notice that even in the most challenging highly non-adiabatic case, only 4 beads are required to converge results, and for the other models above 6 beads, the results are converged. Again, for all cases investigated here, a smaller or equal number of beads is required to converge the SM-NRPMD compared to the previous NRPMD approaches based on the MMST mapping formalism.\cite{ananth2013,chowdhury2017} We have also performed the SM-NRPMD simulations with (i) $\mathrm{s}=\mathrm{Q}$ sampling and dynamics obeying $\bar{\mathrm{s}}=\mathrm{P}$, and (ii) $\mathrm{s}=\mathrm{P}$ sampling and dynamics obeying $\bar{\mathrm{s}}=\mathrm{Q}$. Additional results and discussions are provided in Appendix E.

\begin{figure}[htp]
\includegraphics[width=\columnwidth]{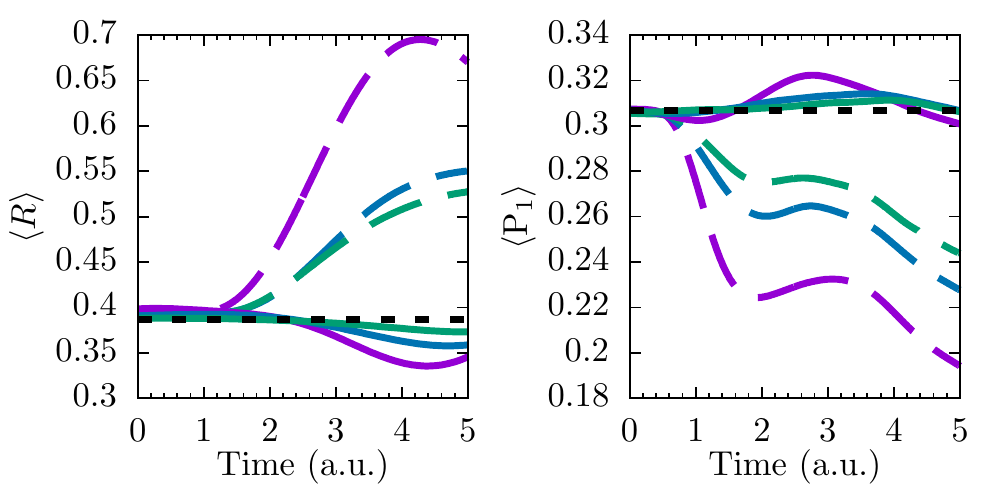}
\caption {Expectation values of the nuclear position operator (left panel) and the electronic population of state 1 (right panel) for model IV (intermediate regime). Results are obtained from SM-NRPMD (solid lines) and NRPMD\cite{richardson2013,richardson2020rev} (dashed lines) with $\mathrm{s}=\mathrm{W}$ results for $N=2$ (magenta), $N=4$ (blue), and $N=6$ (green) beads.}
	\label{fig:crcn}
\end{figure}

Fig. \ref{fig:cnn} presents the nuclear position and the electronic population auto-correlation functions computed from the SM-NRPMD (black solid lines), NRPMD (green dashed lines), as well as numerically exact approach (red dots) for models II, III, and V. Accurately describing electronic Rabi oscillations are essential for non-adiabatic dynamics simulations. Both the SM-NRPMD and the NRPMD agree well with exact results in the adiabatic regime for model II presented in Fig.~\ref{fig:cnn}a, and provide reasonably good results for the model systems in the intermediate regimes presented in Fig.~\ref{fig:cnn}b-c. MV-RPMD\cite{ananth2013} on the other hand, cannot correctly capture the electronic oscillations in these population auto-correlation functions (results not shown), due to the contamination of the true electronic Rabi oscillations with the inter-beads couplings in the mapping ring polymer Hamiltonian.\cite{ananth2013,althorpe2016} 

\begin{figure*}[htp]
\includegraphics[width=\textwidth]{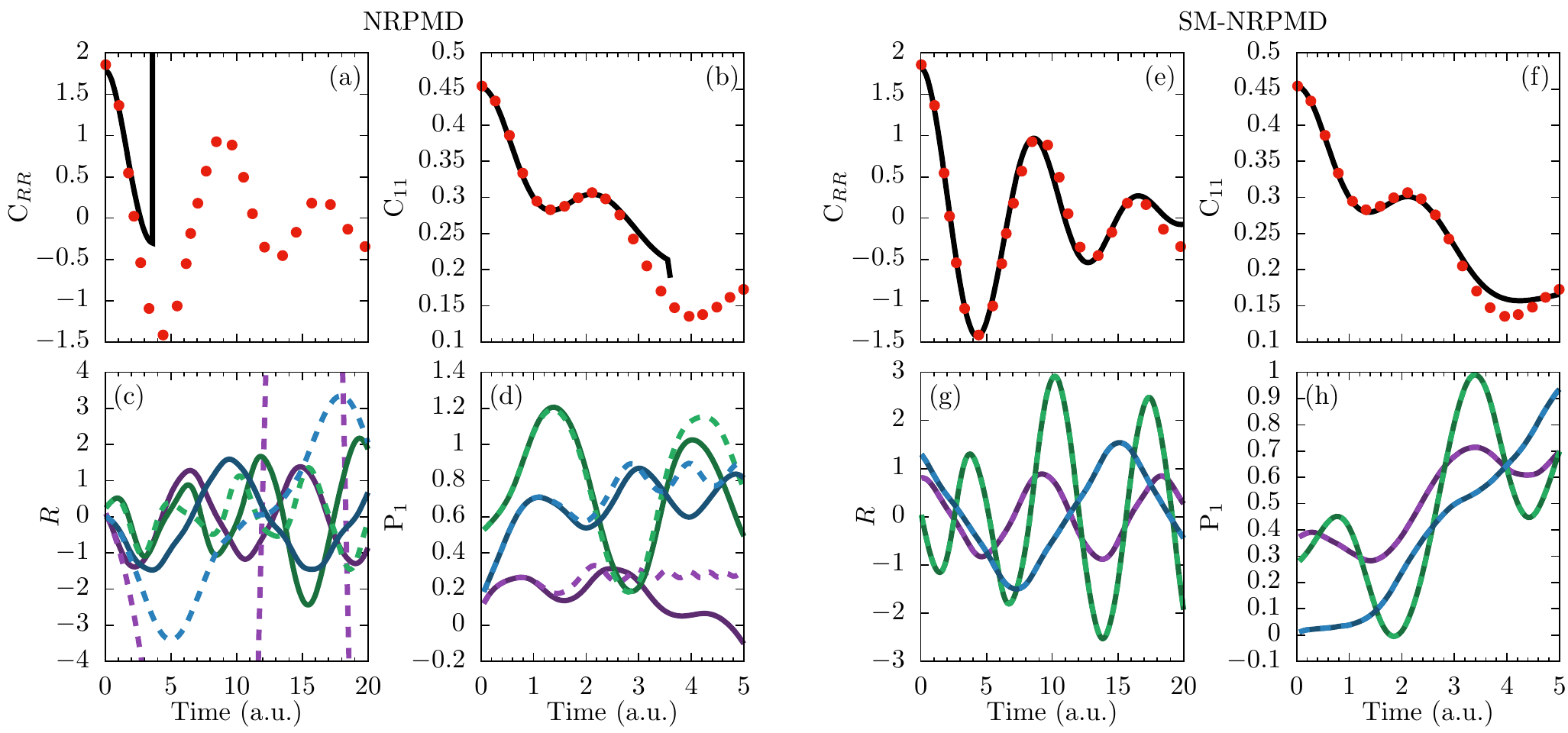}
\caption {The influence of including the quadratic potential $U_{0}(R)$ into the state-dependent Hamiltonian for model V. The left panels (a-d) present the results obtained from the NRPMD method\cite{richardson2013} using the MMST mapping formalism. The Kubo-transformed position auto-TCF (panel a) and population auto-TCF (panel b) are computed with NRPMD (black lines) and exact approach (red dots). The panels (c) and (d) present three representative trajectories (blue, magenta and green) along which $\bar{R}=\frac{1}{N}\sum_{\alpha=1}^{N}R_{\alpha}$ (panel c) and $\bar{\mathrm{P}}_{1}=\frac{1}{N}\sum_{\alpha=1}^{N}\frac{1}{2}\big([q_\alpha]_{1}^2+[p_\alpha]_1^2-1\big)$ are computed (see Eq.~\ref{eq:SC_estimator} in Appendix E). The dynamics are propagated using the NRPMD approach with $U_{0}(R)$ (dashed lines) and without $U_{0}(R)$ (solid lines) inside the state-dependent potential. Right panels (e-h) present the results obtained from the SM-NRPMD method using the spin mapping with $\mathrm{s}=\mathrm{W}$. Panels (g) and (h) provides three representative trajectories along which $\bar{R}=\frac{1}{N}\sum_{\alpha=1}^{N}R_{\alpha}(t)$ (panel c) and $\bar{\mathrm{P}}_{1}=\frac{1}{N}\sum_{\alpha=1}^{N}\frac{1}{2} + r_\mathrm{W}\cdot\cos\theta_\alpha$ are computed using SM-NRPMD.
}
	\label{fig:sm-sep}
\end{figure*}

Fig.~\ref{fig:crcn} presents the time-dependent expectation values of the nuclear position $\langle R\rangle$ (using $\hat{A}=\hat{\mathcal{I}}\otimes\hat{\mathds{1}}_R$ and $\hat{B}=\hat{R}$ in $C^\mathrm{K}_{AB}(t)$), as well as the population for the state 1 $\langle \mathrm{P}_{1}\rangle$ (using $\hat{A}=\hat{\mathcal{I}}\otimes\mathds{1}_R$ and $\hat{B}=|1\rangle\langle 1|$) in model IV (a non-adiabatic case with bias). These expectation values are computed with both SM-NRPMD (solid lines) and NRPMD (dashed lines) and are compared to the exact value. Because the system is under thermal equilibrium, these values should be conserved along the dynamics. As we can see in Fig.~\ref{fig:crcn}, by increasing the number of beads from $N=2$ (magenta), to $N=4$ (blue), and $N=6$ (green), SM-NRPMD (with $\mathrm{s=W}$) almost provides time-independent expectation values. The MMST based approach, such as the NRPMD\cite{richardson2013,richardson2020rev} (dashed lines) can not provide a constant expectation value with the same number of beads. We conjecture that at a large number of beads, SM-NRPMD (with $\mathrm{s}=\mathrm{W}$) might preserve the initial quantum Boltzmann distribution. This conjecture is also corroborate by the numerical evidence that the initial distribution function $\mathrm{Tr_{e}}[{\boldsymbol\Gamma}_{\mathrm{s}}]\cdot e^{-\beta_N\tilde{H}_0({\bf R})}$ (inside Eq.~\ref{SCS-general}) is conserved by the equations of motion in Eqs.~\ref{EOM-SM1}-\ref{EOM-SM3} at the single trajectory level with a large number of beads ($N>32$). To summarize, with a finite number of beads, the SM-NRPMD (with $\mathrm{s}=\mathrm{W}$) largely conserves the initial quantum Boltzmann distribution, providing an almost time-independent expectation value for systems under thermal equilibrium. This is a significant improvement compared to the MMST based NRPMD dynamics.\cite{richardson2013,richardson2017,richardson2020rev}



Compared to the previous NRPMD approach with the MMST formalism, the SM-NRPMD approach provides an additional  advantage that the dynamics is invariant with respect to the splitting between the state-independent potential ${U}_{0}(\hat{R})$ and the state-dependent potential. This is because the spin-mapping formalism explicitly enforces the total population to be 1, such that $[\hat{\mathcal{I}}]_\mathrm{s}(\mathbf{u})=1$. More explicitly, this can be seen in Eq.~\ref{proj11}-\ref{proj22}, leading to  $U_{0}=(\frac{1}{2}+r_\mathrm{s}\cos\theta)\cdot U_{0}+(\frac{1}{2}-r_\mathrm{s}\cos\theta)\cdot U_{0}$. The MMST formalism, on the other hand, does not guarantee this property, and a brief discussion between these two mapping approaches is provided in Appendix A. In order to explicitly demonstrate this advantage of the SM-NRPMD, we incorporate the state-independent quadratic term potential also into the state-dependent Hamiltonian as follows
\begin{align}
\hat{H} = \frac{\hat{P}^2}{2m} + \begin{pmatrix}
\frac{1}{2}m\omega^2\hat{R}^2 + \hat{R} + \epsilon & \Delta \\ 
\Delta & \frac{1}{2}m\omega^2\hat{R}^2 -\hat{R} - \epsilon
\end{pmatrix}.\nonumber
\end{align}

Fig.~\ref{fig:sm-sep} presents the Kubo-transformed nuclear position and population auto-correlation functions for Model V (non-adiabatic case) when including the quadratic potential $U_{0}(R)$ into the state-dependent part. The results are obtained with the NRPMD\cite{richardson2013} using the MMST formalism (panels a-d) and with the SM-NRPMD approach using the spin-mapping formalism (panels e-h). When including $U_0$ into the state-dependent potential, the NRPMD dynamics becomes unstable and completely breaks down at $t\approx3.5$ a.u., as some trajectories within the ensemble start to diverge, causing numerical instabilities. Three representative nuclear position trajectories and population trajectories when including $U_0$ into the state-dependent potential are shown with the dashed lines in Fig.~\ref{fig:sm-sep}c-d, compared to the case when treating the quadratic term $U_0$ as a state-independent potential (solid lines). When individual trajectories have a total population deviated from 1 (as shown in Fig.~\ref{fig:sm-sep}d) in the MMST formalism, the total population also multiplies in front of $U_0$ , resulting in an incorrect force acting on the nuclear DOF, as well as unstable motions. Due to this, including $U_0$ into the state-dependent Hamiltonian could be numerically challenging and eventually causes numerical instabilities. In addition, the results of the auto-correlation functions (before diverging) are different than those obtained in Fig.~\ref{fig:crr}b and Fig.~\ref{fig:cnn}c, indicating that different splitting of state-dependent and state-independent potential in the MMST formalism can lead to different numerical results when using approximate quantum dynamics approaches.\cite{thoss1999} 

Fig.~\ref{fig:sm-sep}e-h present the same comparisons using the spin-mapping approach SM-NRPMD (with $\mathrm{s}=\mathrm{W}$). As expected, the dynamics is invariant under different ways of partitioning $U_{0}(R)$. Fig.~\ref{fig:sm-sep}e and Fig.~\ref{fig:sm-sep}f present the Kubo-transformed TCF when including $U_{0}(R)$ inside the state-dependent potential, providing identical results to those presented in Fig.~\ref{fig:crr}b and Fig.~\ref{fig:cnn}c. In fact, the dynamics is invariant at the single-trajectory level, as clearly indicated in Fig.~\ref{fig:sm-sep}g-h. This is guaranteed because the total population is always bounded by one in spin mapping,\cite{richardson2020} hence the quadratic potential is always $(\frac{1}{2}+r_\mathrm{s}\cos\theta)\cdot U_{0}+(\frac{1}{2}-r_\mathrm{s}\cos\theta)\cdot U_{0}=U_{0}$. This is another unique advantage of using the spin mapping formalism compared to the MMST mapping formalism, in addition to the better preservation of the initial quantum distribution demonstrated in Fig.~\ref{fig:crcn}. Note that in the SM approach, a negative population is still possible in the case of the $\tilde{H}_\mathrm{P}$ and $\tilde{H}_\mathrm{W}$ (see Fig.~\ref{fig:sm-sep}h), but the population is not directly involved in the potential related to $U_0$. In addition, the mapping dynamics of the spin variables $\theta$ and $\varphi$ are bounded on the Bloch sphere of radius $r_\mathrm{s}$, as opposed to un-bounded phase space variables (in the mapping oscillator phase space) in the MMST formalism (see Appendix A). Together, these advantages of the spin-mapping variables make it a more accurate and convenient mapping representation for developing non-adiabatic dynamics methods,\cite{richardson2020, richardson2020rev} and we extend it to the NRPMD dynamics in this work.

\section{Conclusion}
In this paper, we present a new non-adiabatic RPMD method based on the recent development of spin mapping (SM) formalism.\cite{richardson2020} The basis of the spin mapping variables, the spin coherent states, is of the same dimensionality as the electronic Hilbert subspace of the original system. Hence, the SM approach is numerically advantageous compared to the original harmonic oscillator-based mapping approach.\cite{meyer1979,stock1997,thoss1999} These include the total population for a single trajectory is always bounded by one, the dynamics is invariant under different ways of partitioning the state-independent and state-dependent potentials, and the further projections back to the electronic subspace\cite{ananth2010} is not necessary to compute the physical observables. 

Using the spin mapping representation, we derive a general quantum partition function for the coupled electronic-nuclear system, which we refer to as the Spin Coherent State (SCS) Partition Function. We test the performance of the SCS partition function by computing state-dependent nuclear distribution in a two-level system coupled to a harmonic DOF. Our result suggest that the SCS partition function provides the exact quantum results using $N=10$ beads, requiring fewer beads compared to the MMST-based quantum partition functions.\cite{ananth2010,ananth2013,richardson2013,chowdhury2017} Further, the SCS partition function provides an analytical expression of the matrix elements of the thermal Boltzmann operator (Eq.~\ref{eq:Taylorexp}), facilitating the Monte-Carlo numerical simulations. Using various choices of $r_\mathrm{s}$ in the Stratonovich-Weyl transformation, we find that the $\mathrm{s}=\mathrm{Q}$ requires the fewest MC configuration to converge, whereas $\mathrm{s}=\mathrm{W}$ approach requires 10 times more than $\mathrm{s}=\mathrm{Q}$ approach, and $\mathrm{s}=\mathrm{P}$ requires $10^3$ more configurations to converge for $N=10$ beads (this ratio increases when increasing the number of beads). Compared to the MMST based approaches, the $\mathrm{s}=\mathrm{W}$ approach requires a similar amount of configurations and fewer number of beads for convergence compared to the original NRPMD\cite{richardson2013,richardson2017} or CS-RPMD approach\cite{chowdhury2017}.

Using the property of the Stratonovich-Weyl transformation, we further derive the spin-mapping (SM)-NRPMD Hamiltonian, which can be viewed as the unified Hamiltonian of the spin-mapping Hamiltonian and the ring polymer Hamiltonian. Based on this Hamiltonian, we propose the SM-NRPMD dynamics, where the initial sampling is governed by the SCS partition function and the dynamics is governed by the SM-NRPMD Hamiltonian. Using the degrees of freedom of $r_\mathrm{s}$ and $r_{\bar{\mathrm{s}}}$, we find that by choosing $\{\mathrm{s=W},\bar{\mathrm{s}}=\mathrm{W}\}$, SM-NRPMD provides accurate Kubo-transformed nuclear-position auto-correlation function compared to the exact results for model systems that exhibit a broad range of parameters, from electronically adiabatic to the non-adiabatic regime. It can also provide the accurate population auto-correlation function with the correct electronic Rabi oscillation frequency. The accuracy of SM-NRPMD appears to be equivalent (with some slight improvements in certain cases) to those obtained from MMST based non-adiabatic RPMD methods, such as NRPMD\cite{richardson2013,richardson2017} or CS-RPMD,\cite{chowdhury2017} with a similar number of beads to converge the dynamics and a similar amount of trajectories required to converge the calculations.

From our numerical results, the SM-NRPMD seems to preserve the initial quantum Boltzmann distribution by providing a nearly time-independent expectation value of the nuclear position and electronic population. The MMST-based RPMD approaches, on the other hand, failed to generate time-independent expectation value of an observable for systems under thermal equilibrium. Moreover, the SM-NRPMD provides stable and invariant results regardless of how to partition the state-independent and state-dependent potentials, whereas the MMST-based NRPMD dynamics are highly sensitive to the specific choice of splitting the potentials.

To summarize, SM-NRPMD provides accurate electronic non-adiabatic dynamics with explicit nuclear quantization, with additional advantages compared to the original MMST based approaches including a normalized total population along a single trajectory, and the invariant dynamics under different ways of partition of potentials. Future directions include generalizing the current formalism to multi-electronic states,\cite{richardson2020} as well as rigorously derive SM-NRPMD formalism through the recent development of the non-adiabatic Matsubara framework.\cite{chowdhury2021} 

\section*{ACKNOWLEDGMENTS}
This work was supported by the National Science Foundation
CAREER Award under Grant No. CHE-1845747. P.H. appreciates the support from a Cottrell Scholar award (a program by Research Corporation for
Science Advancement). Computing resources were provided by the Center for Integrated Research Computing (CIRC) at the University of Rochester.

\section*{Availability of Data}	
The data that support the findings of this study are available from the corresponding author upon a reasonable request.

\section*{Appendix A: Connection Between the Spin Mapping and the MMST Mapping}
The spin mapping Hamiltonian $H_\mathrm{s}(\mathbf{u})$ in Eq.~\ref{eq:sm-ham} can be transformed into the MMST mapping Hamiltonian. The connection between these two mapping formalisms has been extensively discuss in Ref.~\citenum{richardson2019}. Consider the following variable transformations between the spin mapping variable $\mathbf{u}$ and the MMST mapping variables ${\bf q}=\{q_{1},q_{2}\}$ and ${\bf p}=\{p_{1},p_{2}\}$ as follows
\begin{subequations}\label{trans}
\begin{align}\label{transform}
2r_\mathrm{s}u_x &= {q}_{1}{q}_{2}+{p}_{1}{p}_{2}\\
2r_\mathrm{s}u_y &= {q}_1{p}_2-{q}_2{p}_1\\
2r_\mathrm{s}u_z &= \frac{1}{2}({q}_1^2+{p}_1^2-{q}_2^2-{p}_2^2).
\end{align}
\end{subequations}
Using above transformation in $H_\mathrm{s}(\mathbf{u})$ (Eq.~\ref{eq:sm-ham}) leads to
\begin{align}\label{eq:MMST-sp}
&H_\mathrm{s}= \frac{P^2}{2m}+U_0+\frac{1}{2}(V_1+V_2)\\
&~~~~+\frac{1}{4}(V_1-V_2)\cdot({q}_1^2+{p}_1^2-{q}_2^2-{p}_2^2)+\Delta({q}_1{q}_2+{p}_1{p}_2),\nonumber
\end{align}
which is the MMST Hamiltonian for Hamiltonian $\hat{H}$ in Eq.~\ref{eq:spin_ham}, with a form that separates the trace $\frac{1}{N}\mathrm{Tr_e}\hat{V}$ and trace-less part $\hat{V}-\frac{1}{N}\mathrm{ Tr_e}\hat{V}$, as recommended in the MMST literature.\cite{kelly2012,cotton2013_jcp} 

Using $|\mathbf{u}|^2=\sin^2\theta\cos^2\varphi+\sin^{2}\theta\sin^{2}\varphi+\cos^2\theta=1$ and Eq.~\ref{trans}, one can show that
\begin{equation}
4r_\mathrm{s}={q}_1^2+{q}_2^2+{p}_1^2+{p}_2^2,
\end{equation}
which is often referred to as the total action of the mapping variables.\cite{miller2009} It is also a conserved quantity of the MMST Hamiltonian in Eq.~\ref{eq:MMST-sp}. Using this property, one can rewrite the MMST Hamiltonian in Eq.~\ref{eq:MMST-sp} as follows
\begin{align}\label{eq:MMST}
&H_\mathrm{s}=\frac{P^2}{2m}+U_0+\Delta({q}_1{q}_2+{p}_1{p}_2)\\
&+\frac{1}{2}V_1\cdot(\frac{{q}_1^2+{p}_1^2-{q}_2^2-{p}_2^2}{2}+1+2r_\mathrm{s}-2r_\mathrm{s}) \nonumber\\
&+\frac{1}{2}V_2\cdot(\frac{{q}_2^2+{p}_2^2-{q}_1^2-{p}_1^2}{2}+1+2r_\mathrm{s}-2r_\mathrm{s}) \nonumber\\
&=\frac{P^2}{2m}+U_0+\Delta({q}_1{q}_2+{p}_1{p}_2)+\frac{1}{2}\sum_{n=1}^{2}V_n\cdot({q}_n^2+{p}_n^2-\gamma),\nonumber
\end{align}
where the MMST mapping oscillators' zero-point energy correction is defined as
\begin{equation}
\gamma=2r_\mathrm{s}-1.
\end{equation}
Connecting to the spin-mapping Hamiltonian, we can hence identify $H_\mathrm{Q}$ as MMST formalism with $\gamma=0$, $H_\mathrm{P}$ as $\gamma=2$ and $H_\mathrm{W}$ as $\gamma=\sqrt{3}-1$ (which is the recommended value in symmetric quasi-classical (SQC) approach\cite{cotton2013_jcp} that was derived based on analogy with spin). 

Note that the MMST Hamiltonian in Eq.~\ref{eq:MMST} has been historically introduced through the mapping relation $|i\rangle\langle j|\to {a}^{\dagger}_{i}\hat{a}_{j}=\frac{1}{\sqrt{2}}(\hat{q}_i - i\hat{p}_i)\cdot\frac{1}{\sqrt{2}}(\hat{q}_j + \hat{p}_j)$, as well as using $[p_{i},q_{i}]=i$ (or effectively,  $[p_{i},q_{i}]=i\gamma$ for the adjusted mapping oscillator Zero-Point Energy), hence 
\begin{align}\label{MMST-general}
&\sum_{ij}V_{ij}(\hat{R})|i\rangle\langle j|\to\sum_{ij}V_{ij}{a}^{\dagger}_{i}\hat{a}_{j}\\
&=\frac{1}{2}\sum_{ij}{V}_{ij}(\hat{R})(\hat{q}_i\hat{q}_j+ \hat{p}_i \hat{p}_j - \gamma\delta_{ij}).\nonumber
\end{align}
The fundamental differences between the spin mapping Hamiltonian in Eq.~\ref{eq:spin_ham} and the MMST Hamiltonian in Eq.~\ref{eq:MMST} are (i) For a two-state system, the spin mapping Hamiltonian only has two independent variables $\theta$ and $\varphi$, thus, the same dimensionality of the original electronic subspace, whereas the MMST Hamiltonian has effectively four independent variables $\{p_1, q_1, p_2, q_2\}$ hence a larger dimensionality. (ii) The total population of the spin mapping is always bounded by 1, whereas this is not always guaranteed for the MMST mapping formalism.\cite{richardson2019} 

\section*{Appendix B: Equations of motion for $\theta$ and $\varphi$}
Eq.~\ref{u-dyn} can also be equivalently expressed as EOMs in $\theta$ and $\varphi$. Using $\dot{u}_{z}=-\dot{\theta}\sin\theta=H_{x}u_{y}-H_{y}u_{x}$ as well as $\dot{u}_{x}=\dot{\theta}\cos\theta\cos\varphi-\dot{\varphi}\sin\theta\sin\varphi=H_{y}u_{z}-H_{z}u_{y}$ we can derive the following equations 
\begin{subequations}
\begin{align}
\dot{\theta} = &\frac{1}{\hbar}(- H_x\sin\varphi+H_y\cos\varphi), \\
\dot{\varphi} = & \frac{1}{\hbar}\left(H_z - H_x\frac{\cos\varphi}{\tan\theta} - H_y\frac{\sin\varphi}{\tan\theta}\right).
\end{align}
\label{eq:eom}
\end{subequations}

It is interesting to note that the above equations are equivalent to the following
\begin{subequations}\label{eq:eomtp}
\begin{align}
\dot{\theta} = &\frac{1}{r_\mathrm{s}\sin\theta}\frac{\partial {H}_\mathrm{s}(\mathbf{u})}{\partial\varphi} \\
\dot{\varphi} = &-\frac{1}{r_\mathrm{s}\sin\theta}\frac{\partial{H}_\mathrm{s}(\mathbf{u})}{\partial\theta}
\end{align}
\end{subequations}
from which we obtain the conjugate variables $\dot{\varphi}$ and $r_\mathrm{s}\cos\theta$ related to the spin mapping representation, where the latter plays the role of conjugate momentum\cite{klauder1979} to $\varphi$ as
\begin{subequations}
\begin{align}
\frac{d}{dt}(r_\mathrm{s}\cos\theta) &=-\frac{\partial {H}_\mathrm{s}(\bf u)}{\partial\varphi} \\
\dot{\varphi} & =\frac{\partial {H}_\mathrm{s}(\bf u)}{\partial (r_\mathrm{s}\cos\theta)}.
\end{align}
\end{subequations}
The relationship between the Hamiltonian ${H}_\mathrm{s}(\bf u)$ and Lagrangian is $H(\theta,\varphi) = \dot{\varphi}\cdot(r_\mathrm{s}\cos\theta) - {\cal{L}}(\varphi,\dot{\varphi})$.

Note that under the non-equilibrium condition with focused initial condition, such as $[|1\rangle\langle 1|]_\mathrm{s}(\mathbf{u}) =\frac{1}{2}+r_\mathrm{s}\cos\theta=1$ (Eq.~\ref{proj11}), it requires $\cos\theta=1$ under $\mathrm{s}=\mathrm{Q}$, which makes the above EOM ill-defined in terms of $1/\sin\theta$ and $1/\tan\theta$. Thus, $\bf u$ is a more convenient dynamical variable than $\{\theta, \phi\}$ for this scenario. Under the thermal equilibrium condition (such as examples in this paper), the system will never reach to $\theta=0$, we find that using Eq.~\ref{eq:eomtp} is numerically more convenient. We hence use the velocity Verlet algorithm to evolve $\theta$ and $\varphi$, which avoids the necessity to compute any derivative of the potential, as is the case for $\dot{\mathbf{u}}$.

\section*{Appendix C: Elementary Relations in Spin Mapping Representation}
Here, we verify several basic properties of the Stratonovich-Weyl s-transforms. We begin by explicitly expressing $\hat{w}_\mathrm{s}$ defined as follows
\begin{align}
\hat{w}_\mathrm{s} (\mathbf{u})= \begin{pmatrix}
\frac{1}{2}+r_\mathrm{s}\cos\theta & r_\mathrm{s} \sin\theta \cdot e^{-i\varphi}\\ 
r_\mathrm{s} \sin\theta \cdot e^{i\varphi} & \frac{1}{2}-r_\mathrm{s}\cos\theta
\end{pmatrix}\equiv\begin{pmatrix}
w^\mathrm{s}_{11} & w^\mathrm{s}_{12}\\ 
w^\mathrm{s}_{21} & w^\mathrm{s}_{22}
\end{pmatrix}.\nonumber
\end{align}
First, we verify that $\int\mathrm{d}\mathbf{u}A_\mathrm{s}(\mathbf{u})=\mathrm{Tr}_\mathrm{e}[\hat{A}]$ by computing $\mathrm{Tr}_\mathrm{e}[\hat{A}]$ for a general operator $\hat{A}$ as follows
\begin{align}
&\int\mathrm{d}\mathbf{u}A_\mathrm{s}(\mathbf{u})=\frac{1}{2\pi}\int_0^{2\pi}\mathrm{d}\varphi\int_0^\pi\mathrm{d}\theta\sin\theta\mathrm{Tr}_\mathrm{e}[\hat{A}\hat{w}_{\mathrm{s}}] \\
&=\frac{1}{2\pi}\int_0^{2\pi}\mathrm{d}\varphi\int_0^\pi\mathrm{d}\theta\sin\theta \nonumber\\
&~~~\times\Big(A_{11}(\frac{1}{2}+r_{\mathrm{s}}\cos\theta)+A_{12}r_{\mathrm{s}}(\sin\theta\cos\varphi+i\sin\theta\sin\varphi) \nonumber\\
&~~~~~~+A_{21}r_{\mathrm{s}}(\sin\theta\cos\varphi-i\sin\theta\sin\varphi)+A_{22}(\frac{1}{2}-r_{\mathrm{s}}\cos\theta)\Big)\nonumber\\
&=A_{11}+A_{22}=\mathrm{Tr}_\mathrm{e}[\hat{A}],\nonumber
\end{align}
where we have used the elementary results of integrals $\int_0^\pi\mathrm{d}\theta\sin\theta = 2$, $\int_0^\pi\mathrm{d}\theta\cos\theta\sin\theta = 0$,  $\int_0^{2\pi}\mathrm{d}\varphi\cos\varphi = 0$,
and $\int_0^{2\pi}\mathrm{d}\varphi\sin\varphi = 0$. Using these integrals, it is also straightforward to verify that 
\begin{equation}
\int\mathrm{d}\mathbf{u}\hat{w}_{\mathrm{s}}(\mathbf{u}) = \frac{1}{2\pi}\int_0^{2\pi}\mathrm{d}\varphi\int_0^\pi\mathrm{d}\theta\sin\theta\big(\frac{1}{2}\hat{{\mathcal{I}}}+r_\mathrm{s}\mathbf{u}\cdot\hat{\boldsymbol{\sigma}}\big) =\hat{{\mathcal{I}}},
\end{equation}
proving the resolution of identity in the spin mapping coherent state basis.

For the Stratonovich-Weyl transform of the product of two operators $\hat{A}$ and $\hat{B}$, one can show that
\begin{align}\label{TrAB}
&\int\mathrm{d}\mathbf{u}A_\mathrm{s}B_{\bar{\mathrm{s}}}(\mathbf{u}) = \int\mathrm{d}\mathbf{u}\mathrm{Tr}_\mathrm{e}[\hat{A}\hat{w}_{\mathrm{s}}]\mathrm{Tr}_\mathrm{e}[\hat{B}\hat{w}_{\bar{\mathrm{s}}}]\\
&=\int\mathrm{d}\mathbf{u} (A_{11}w^\mathrm{s}_{11}+A_{12}w^\mathrm{s}_{21}+A_{21}w^\mathrm{s}_{12}+A_{22}w^\mathrm{s}_{22})\nonumber\\
&~~~~~\cdot(B_{11}w^{\bar{\mathrm{s}}}_{11}+B_{12}w^{\bar{\mathrm{s}}}_{21}+B_{21}w^{\bar{\mathrm{s}}}_{12}+B_{22}w^{\bar{\mathrm{s}}}_{22}).\nonumber
\end{align}
Note that any of the above terms that contains $\int_{0}^{2\pi}d\varphi e^{\pm i\varphi}$ or $\int_{0}^{2\pi}d\varphi e^{\pm 2i\varphi}$ will be zero. Hence, only the terms without $e^{\pm i\varphi}$ survive. They are either $w^\mathrm{s}_{11}w^{\bar{\mathrm{s}}}_{11}$, $w^\mathrm{s}_{22}w^{\bar{\mathrm{s}}}_{22}$, or $w^\mathrm{s}_{12}w^{\bar{\mathrm{s}}}_{21}$, $w^\mathrm{s}_{21}w^{\bar{\mathrm{s}}}_{12}$. For the term related to $A_{11}B_{11}\cdot\int\mathrm{d}\mathbf{u}w^\mathrm{s}_{11}w^{\bar{\mathrm{s}}}_{11}$, the integral related to the mapping variables is
\begin{equation}
\int\mathrm{d}\mathbf{u}w^\mathrm{s}_{11}w^{\bar{\mathrm{s}}}_{11}=\int_0^\pi\mathrm{d}\theta\sin\theta(\frac{1}{2}+r_\mathrm{s}\cos\theta)(\frac{1}{2}+r_{\bar{\mathrm{s}}}\cos\theta)=1\nonumber
\end{equation}
where we used the fact that $r_\mathrm{s}\cdot r_{\bar{\mathrm{s}}}=3/4$, $\int_0^\pi\mathrm{d}\theta\sin\theta = 2$, and $\int_0^\pi\mathrm{d}\theta\sin\theta\cos^2\theta = \frac{2}{3}$. Similarly, one can show that $\int\mathrm{d}\mathbf{u}w^\mathrm{s}_{22}w^{\bar{\mathrm{s}}}_{22}=1$ as well. The other non-zero terms are $A_{12}B_{21}\cdot\int\mathrm{d}\mathbf{u}w^\mathrm{s}_{21}w^{\bar{\mathrm{s}}}_{12}$ and $A_{21}B_{12}\cdot\int\mathrm{d}\mathbf{u}w^\mathrm{s}_{12}w^{\bar{\mathrm{s}}}_{21}$, with the weighting factor
\begin{equation}
\int\mathrm{d}\mathbf{u}w^\mathrm{s}_{21}w^{\bar{\mathrm{s}}}_{12}=\int\mathrm{d}\mathbf{u}w^\mathrm{s}_{12}w^{\bar{\mathrm{s}}}_{21}=\int_0^\pi\mathrm{d}\theta\sin\theta \cdot r_\mathrm{s}r_{\bar{\mathrm{s}}}\cdot\sin^2\theta=1,\nonumber
\end{equation}
where $\int_0^\pi\mathrm{d}\theta\sin^3\theta=4/3$. Putting all of these together, we have
\begin{align}
\int\mathrm{d}\mathbf{u}A_\mathrm{s}B_{\bar{\mathrm{s}}}(\mathbf{u})=&A_{11}B_{11}++A_{12}B_{21}+A_{21}B_{12}+A_{22}B_{22} \nonumber\\
=&\mathrm{Tr}_{\mathrm{e}}[\hat{A}\hat{B}],
\end{align}
which is Eq.~\ref{eq:QPQ} of the main text.

\section*{Appendix D: Derivation of the SCS partition function}
We derive an analytic expression of the Boltzmann operator in the spin mapping representation. For that we first Taylor expand it as
\begin{align}\label{boltz-taly}
e^{-\beta_N\frac{1}{\hbar}\mathbf{H}_\alpha\cdot\hat{\mathbf{S}}} = & \hat{\mathcal{I}}-\beta_N\frac{1}{\hbar}\mathbf{H}_\alpha\cdot\hat{\mathbf{S}}+\frac{\beta_N^2}{2!}(\frac{1}{\hbar}\mathbf{H}_\alpha\cdot\hat{\mathbf{S}})^2 \nonumber\\
&-\frac{\beta_N^3}{3!}(\frac{1}{\hbar}\mathbf{H}_\alpha\cdot\hat{\mathbf{S}})^3+\cdots.
\end{align}
Using the fact that $(\frac{1}{\hbar}\mathbf{H}_\alpha\cdot\hat{\mathbf{S}})^2=|\mathbf{H}_\alpha|^2/4\times\hat{\mathcal{I}}$, Eq,~\ref{boltz-taly} leads to an expression with two different types of terms that can be identified as Taylor expansions of hyperbolic cosine and hyperbolic sine as follows
\begin{align}
e^{-\beta_N\frac{1}{\hbar}\mathbf{H}_\alpha\cdot\hat{\mathbf{S}}}=&  \sum_{j=0}\frac{1}{(2j)!}\Big(\beta_N\frac{|\mathbf{H}_\alpha|}{2}\Big)^{2j}\hat{\mathcal{I}}-\frac{2\mathbf{H}_\alpha\cdot\hat{\mathbf{S}}}{\hbar|\mathbf{H}_\alpha|} \\
&\times\sum_{j=0}\frac{1}{(2j+1)!}\Big(\beta_N\frac{|\mathbf{H}(R_\alpha)|}{2}\Big)^{2j+1} \nonumber\\
=& \cosh\frac{\beta_N|\mathbf{H}_\alpha|}{2}\hat{\mathcal{I}}-\frac{\mathbf{H}_\alpha\cdot\hat{\boldsymbol{\sigma}}}{|\mathbf{H}_\alpha|}\sinh\frac{\beta_N|\mathbf{H}_\alpha|}{2}.\nonumber
\end{align}
Using the above result as well as the identity $(\mathbf{A}\cdot\hat{\boldsymbol{\sigma}})(\mathbf{B}\cdot\hat{\boldsymbol{\sigma}}) = \mathbf{A}\cdot\mathbf{B}\hat{\mathcal{I}}+i\mathbf{A}\wedge\mathbf{B}\cdot\hat{\boldsymbol{\sigma}}$, one can show that
\begin{align}
&e^{-\beta_N\frac{1}{\hbar}\mathbf{H}_\alpha\cdot\hat{\mathbf{S}}}\hat{w}_\mathrm{s}(\mathbf{u}_\alpha) \\
&=\cosh\big(\frac{\beta_N|\mathbf{H}_\alpha|}{2}\big)\big(\frac{1}{2}\hat{\mathcal{I}}+r_\mathrm{s}\mathbf{\mathbf{u}}_\alpha\cdot\hat{\boldsymbol{\sigma}}\big)-\frac{1}{|\mathbf{H}_\alpha|} \sinh\big(\frac{\beta_N|\mathbf{H}_\alpha|}{2}\big)\nonumber\\
&~~~\times\Big(\frac{1}{2}\mathbf{H}_\alpha\cdot\hat{\boldsymbol{\sigma}}+r_\mathrm{s}(\mathbf{H}_\alpha\cdot\mathbf{u}_\alpha\hat{\mathcal{I}}+i\mathbf{H}_\alpha\wedge\mathbf{u}_\alpha\cdot\hat{\boldsymbol{\sigma}})\Big) \nonumber\\
&=\Big(\frac{1}{2}\cosh\frac{\beta_N|\mathbf{H}_\alpha|}{2}-r_\mathrm{s}\frac{\mathbf{H}_\alpha}{|\mathbf{H}_\alpha|}\cdot\mathbf{u}_\alpha\sinh\frac{\beta_N|\mathbf{H}_\alpha|}{2}\Big)\hat{\mathcal{I}} \nonumber\\
&~~~+\Big(r_\mathrm{s}\mathbf{u}_\alpha\cosh\frac{\beta_N|\mathbf{H}_\alpha|}{2}-\frac{1}{|\mathbf{H}_\alpha|}\big(\frac{\mathbf{H}_\alpha}{2}+ir_\mathrm{s}\mathbf{H}_\alpha\wedge\mathbf{u}_\alpha\big) \nonumber\\
&~~~~~~\times\sinh\frac{\beta_N|\mathbf{H}_\alpha|}{2}\Big)\cdot\hat{\boldsymbol{\sigma}}. \nonumber
\end{align}
This is the general expression of the $\boldsymbol{\Gamma}_\mathrm{s}$ in the SCS partition function (Eq.~\ref{SCS-general}) for any SW transformation. 

\section*{Appendix E: NRPMD and Meanfield-RPMD}
The NRPMD method\cite{richardson2013} was first proposed by Thoss and Richardson as a model dynamics. Recently, it was rigorously derived from the non-adiabatic Matsubara dynamics formalism\cite{chowdhury2021}. It uses the MMST formalism (Eq.~\ref{MMST-general}) to describe the electronic DOFs and the ring polymer path-integral formalism to describe the nuclear DOFs. When operators $\hat{A}$ and $\hat{B}$  are both functions of $\hat{R}$, the NRPMD TCF is expressed as 
\begin{align}\label{eqn:NRPMD}
C_{AB}(t) &=\frac{1}{\cal{Z}}\lim_{N\rightarrow\infty} \int \mathrm{d}\{R_\alpha\} \int \mathrm{d}\{P_\alpha\}\int \mathrm{d}\{{\bf q}_\alpha\} \int \mathrm{d}\{{\bf p}_\alpha\}\nonumber\\
&\times \mathrm{Tr_{e}}[{\boldsymbol\Gamma}'({\bf R},\mathbf{q},\mathbf{p})]e^{-\beta_N H_\mathrm{rp}({\bf R})} \bar{A}(\mathbf{R}) \bar{B}(\mathbf{R}_t),
\end{align}
where $H_\mathrm{rp}({\bf R})= \sum_{\alpha=1}^{N}\frac{P_\alpha^2}{2m} + \frac{m}{2\beta_N^2\hbar^2}(R_{\alpha}- R_{\alpha-1})^{2} + U_{0}(R_{\alpha})$ corresponds to the ring polymer Hamiltonian with the state-independent potential, $\bar{A}({\bf R})=\frac{1}{N}\sum_{\alpha=1}^{N}A(R_{\alpha})$ and $\bar{B}({\bf R}_t)=\frac{1}{N}\sum_{\alpha=1}^{N}B(R_{\alpha}(t))$, and ${\boldsymbol\Gamma}'({\bf R},\mathbf{q},\mathbf{p})$ is expressed as\cite{richardson2013}
\begin{equation}
{\boldsymbol\Gamma}'({\bf R},\mathbf{q},\mathbf{p})= e^{-{\mathcal{G}_N}}\prod_{\alpha=1}^{N}[{\pmb{\mathcal M}}'(R_\alpha){\bf q}_{\alpha}{\bf q}^\mathrm{T}_{\alpha}{\pmb{\mathcal M}}'(R_\alpha){\bf p}_{\alpha}{\bf p}^\mathrm{T}_{\alpha}],\nonumber
\end{equation}
with $\mathcal{G}_N=\sum_{\alpha=1}^{N}(\mathbf{q}_\alpha^{\mathrm{T}}\mathbf{q}_\alpha+\mathbf{p}_\alpha^{\mathrm{T}}\mathbf{p}_\alpha)$, and $\mathcal{M}'_{ij}(R_\alpha) = \langle i |e^{-\frac{1}{2}\beta_N \hat{V}_\mathrm{e}(R_\alpha)}|j \rangle$. Note that $\mathrm{Tr_{e}}[{\boldsymbol\Gamma}'({\bf R},\mathbf{q},\mathbf{p})]$ can also be equivalently expressed as   $\mathrm{Tr_{e}}[{\boldsymbol\Gamma}'({\bf R},\mathbf{q},\mathbf{p})]= e^{-\mathcal{G}_N}\prod_{\alpha=1}^{N} [{\bf p}^\mathrm{T}_{\alpha-1}\pmb{\mathcal M}'(R_\alpha){\bf q}_{\alpha}]\cdot \big[{\bf q}_{\alpha}^{\mathrm{T}}\pmb{\mathcal{M}}'(R_\alpha){\bf p}_\alpha\big]$. 

The dynamics is governed by the following NRPMD Hamiltonian\cite{richardson2013}
\begin{align}\label{eqn:HNRPMD}
&H_{N}=
\frac{1}{N}\sum_{\alpha=1}^{N}\Big[\frac{{P}^2_{\alpha}}{2m}+\frac{m}{2\beta^{2}_{N}\hbar^{2}}(R_{\alpha}-R_{\alpha-1})^{2}+U_{0}(R_\alpha)\nonumber\\
&+\frac{1}{2}\sum_{i,j=1}^{\mathcal{K}}{V}_{ij}(R_\alpha)\Big([{\bf p}_{\alpha}]_i [{\bf p}_{\alpha}]_j + [{\bf q}_{\alpha}]_i [{\bf q}_{\alpha}]_j -\delta_{ij}\Big)\Big].
\end{align}
The NRPMD Hamiltonian was derived from both the partition function expression\cite{chowdhury2019} as well as from a quantum Liouvillian.\cite{chowdhury2021} It is closely related to the SM-NRPMD Hamiltonian in Eq.~\ref{SW-NRP} through the transformation in Eq.~\ref{transform}.

When $\hat{A}=|i\rangle\langle i|$ and $\hat{B}=|j\rangle\langle j|$, the NRPMD TCF is 
\begin{align}
C_{AB}(t) =&\frac{1}{\cal{Z}}\lim_{N\rightarrow\infty} \int \mathrm{d}\{R_\alpha\} \int \mathrm{d}\{P_\alpha\}\int \mathrm{d}\{{\bf q}_\alpha\} \int \mathrm{d}\{{\bf p}_\alpha\}\nonumber\\
&\times \mathrm{Tr_{e}}[{\boldsymbol\Gamma}'|i\rangle\langle i|]e^{-\beta_{N} H_\mathrm{rp}({\bf R})} [\hat{a}^{\dagger}_{j}\hat{a}_{j}]_{N}(t),
\end{align}
with the electronic state estimator\cite{richardson2013,richardson2017}
\begin{equation}\label{eq:SC_estimator}
P_{j}(t)=[\hat{a}^{\dagger}_{j}\hat{a}_{j}]_{N}=\frac{1}{N}\sum_{\alpha=1}^{N}\frac{1}{2}\big([{\bf q}_\alpha]_{j}^2+[{\bf p}_{\alpha}]_{j}^2-1\big).
\end{equation}

The mean-field (MF)-RPMD approach\cite{hele2011} can be viewed as a special limit of the NRPMD TCF Eq.~\ref{eqn:NRPMD} by analytically integrating out the mapping variables in Eq.~\ref{eqn:NRPMD} at $t=0$. The MF-RPMD TCF\cite{hele2011,ananth2013} is
\begin{align}\label{eqn:MFRPMD}
C_{AB}^{\mathrm{MF}}(t) &=\frac{1}{\cal{Z}}\lim_{N\rightarrow\infty} \int \mathrm{d}\{R_\alpha\} \int \mathrm{d}\{P_\alpha\}\\
&\times \mathrm{Tr}_{\mathrm{e}}[{\bf \Gamma''}({\bf R})] e^{-\beta_N {H_{\mathrm{rp}}}}\bar{A}({\bf R})\bar{B}({\bf R}_t),\nonumber
\end{align}
where ${\bf \Gamma''}({\bf R})=\prod_{\alpha=1}^{N}\pmb{\mathcal{M}}(R_\alpha)$, and $\mathcal{M}_{ij}(R_\alpha) = \langle i |e^{-\beta_N \hat{V}_\mathrm{e}(R_\alpha)}|j \rangle$. The MF-RPMD dynamics is governed by the MF-RPMD effective Hamiltonian\cite{hele2011,ananth2013} $H_\mathrm{MF}=H_{\mathrm{rp}}-\frac{1}{\beta}\ln |\mathrm{Tr}_{\mathrm{e}}[{\bf \Gamma}''({\bf R})]|$.
Note that MF-RPMD is not a new method and has been derived without using mapping representation.\cite{hele2011} 

\section*{Appendix F: Additional results of SM-NRPMD}
In this section, we explore other possible choices of $\{\mathrm{s}, \bar{\mathrm{s}}\}$ in the SM-NRPMD dynamics. Using the model systems, we find that the $\{\mathrm{s=Q},\bar{\mathrm{s}}=\mathrm{P}\}$ choice provides the most efficient initial sampling, which requires 10 times smaller configurations (trajectories) than the $\{\mathrm{s=W},\bar{\mathrm{s}}=\mathrm{W}\}$ choice when using $N=6$ beads for a converged dynamics (due to a more severe sign problem). Thus, for pure quantum statistical quantities, $\{\mathrm{s=Q}\}$ provides the most efficient sampling. The TCF dynamics, unfortunately, seems to require more beads to converge. For correlation function calculation, this disadvantage counterbalances its advantage and makes the $\{\mathrm{s=W},\bar{\mathrm{s}}=\mathrm{W}\}$ more favorable. 
\begin{figure}[htp]
	\includegraphics[width=\columnwidth]{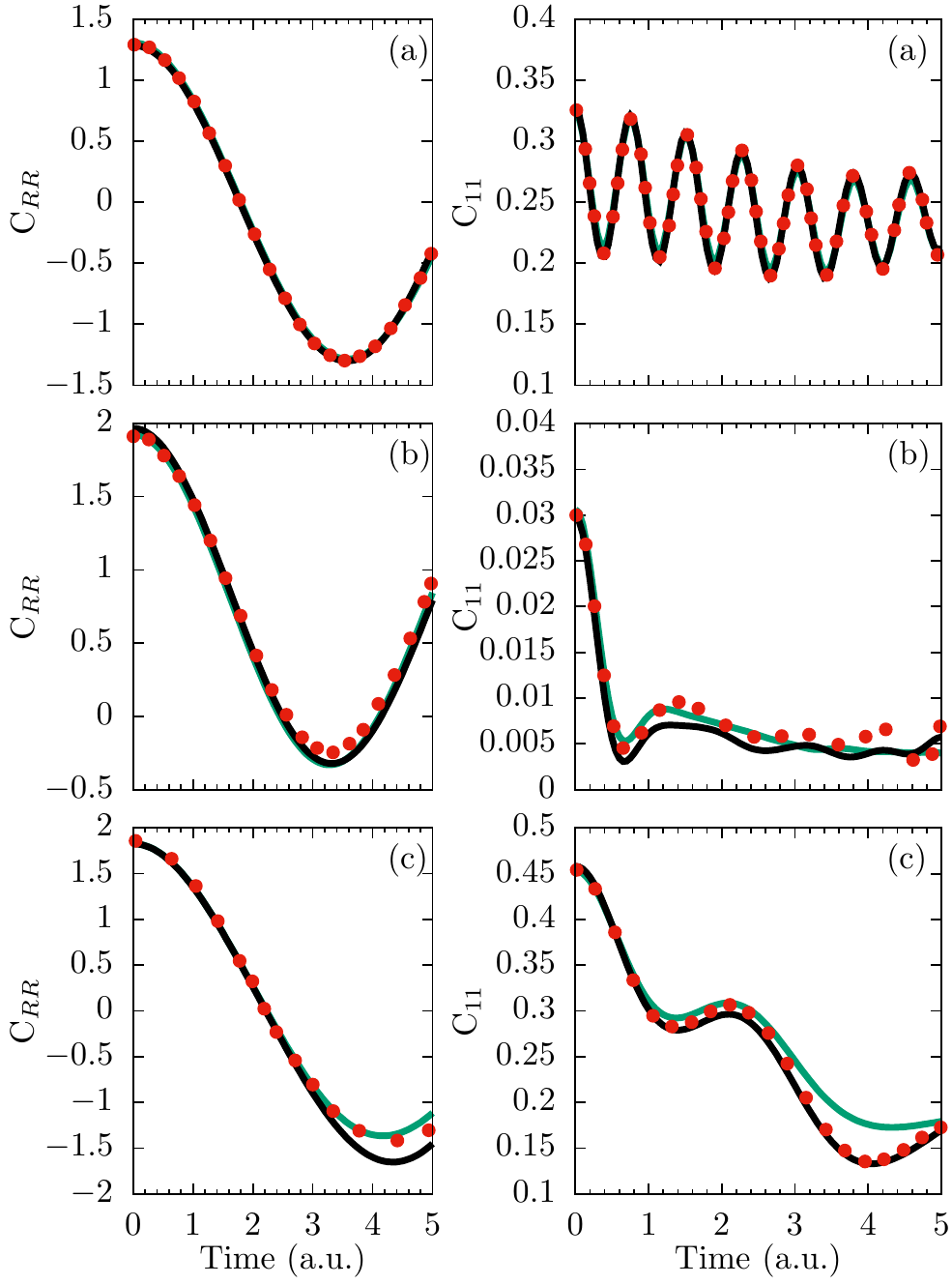}
    \caption {Position auto-correlation functions (left panels) and its corresponding population auto-correlation functions (right panels) with $\mathrm{s}=\mathrm{P}$ in black solid lines of models II (4 beads), III (6 beads), V (4 beads) respectively in panels (a), (b), (c). Similar calculations with $\mathrm{s}=\mathrm{Q}$ in green solid lines for models  II (8 beads), III (18 beads), V (18 beads). SM-NRPMD is compared to the exact result (red dots).}
    \label{fig6}
\end{figure}

The choice of $\{\mathrm{s}=\mathrm{P},\bar{\mathrm{s}}=\mathrm{Q}\}$, on the other hand, provides a more accurate electronic auto-correlation function at a longer time. This finding agrees with the out of equilibrium calculations, which conclude that the $\bar{\mathrm{s}}=\mathrm{Q}$ choice in the mapping Hamiltonian provides the most accurate electronic dynamics.\cite{richardson2019} For the thermal TCF calculation, however, the initial sampling with the choice of $\mathrm{s}=\mathrm{P}$ typically requires 10 to $10^2$ more configurations to achieve a numerical convergence. 

Fig.~\ref{fig6} presents the SM-NRPMD position and population auto-correlation functions for models II, III and V with the choice of $\{\mathrm{s=Q},\bar{\mathrm{s}}=\mathrm{P}\}$ (green solid lines) as well as $\{\mathrm{s=P},\bar{\mathrm{s}}=\mathrm{Q}\}$ (black solid lines), compared to the numerically exact results (red dots). Indeed, the $\mathrm{s}=\mathrm{P}$ choice, hence $\bar{\mathrm{s}}=\mathrm{Q}$ for the SM-NRPMD Hamiltonian provides the most accurate electronic dynamics (more accurate than the $\{\mathrm{s=W},\bar{\mathrm{s}}=\mathrm{W}\}$ results in Fig.~\ref{fig:cnn}), with the price of using more trajectories to achieve numerical convergence of the TCF.

The $\{\mathrm{s=Q},\bar{\mathrm{s}}=\mathrm{P}\}$ calculations (green) show a generally good agreement with the exact results, but require more beads to converge (with up to 18 beads in the model calculations presented here). Calculations with such a large number of beads are made possible by the fast convergence of $\mathrm{s}=\mathrm{Q}$ sampling (with $10^6$ configurations for $N=18$ beads).
\begin{figure}[htp]
\includegraphics[width=\columnwidth]{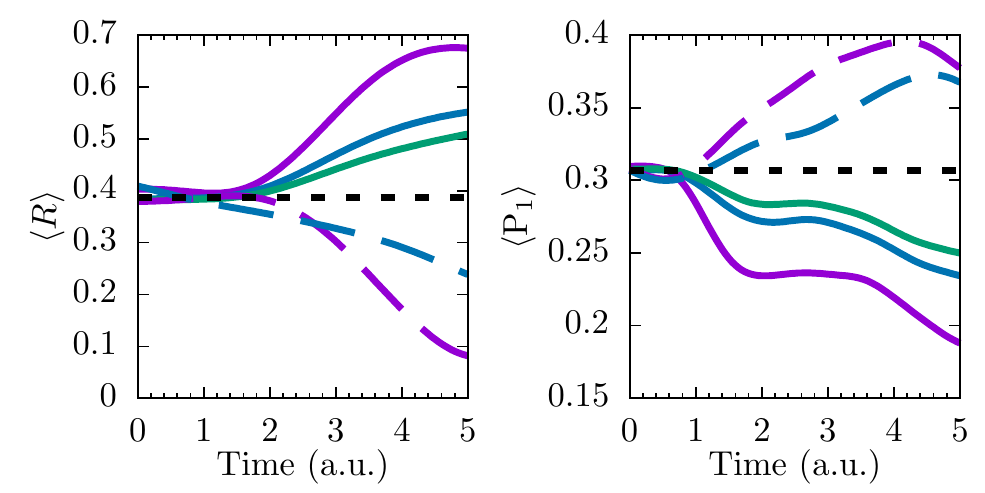}
\caption {Expectation values of the nuclear position operator (left panel) and of the electronic population of state 1 (right panel) for model IV (intermediate regime). Results are obtained from SM-NRPMD using $\{\mathrm{s=Q},\bar{\mathrm{s}}=\mathrm{P}\}$ (solid lines) and $\{\mathrm{s=P},\bar{\mathrm{s}}=\mathrm{Q}\}$ (dashed lines), with $N=2$ (magenta), $N=4$ (blue) and $N=6$ (green) beads, respectively. Numerically exact results are shown in black dotted lines.}
\label{fig7}
\end{figure}

Fig.~\ref{fig7} presents the expectation values of the nuclear position operator and electronic population of state 1 for $\{\mathrm{s=Q},\bar{\mathrm{s}}=\mathrm{P}\}$ (solid lines) and $\{\mathrm{s=P},\bar{\mathrm{s}}=\mathrm{Q}\}$ (dashed lines). Both choices failed to provide the time-independent expectation values, in contrast to the case of $\{\mathrm{s=W},\bar{\mathrm{s}}=\mathrm{W}\}$ results presented in Fig.~\ref{fig:crcn}. It also seems that both $\{\mathrm{s=Q},\bar{\mathrm{s}}=\mathrm{P}\}$ and $\{\mathrm{s=P},\bar{\mathrm{s}}=\mathrm{Q}\}$ require even more beads to converge these expectation values compared to the MMST-based NRPMD method\cite{richardson2020rev} as shown in Fig.~\ref{fig:crcn} (dashed lines).

%
%

%
%
\end{document}